# Development of different methods and their efficiencies for the estimation of diffusion coefficients following the diffusion couple technique

Varun A. Baheti [1,2] and Aloke Paul [1,*]

[1] Department of Materials Engineering, Indian Institute of Science, Bangalore 560012, India
[2] Department of Metallurgical and Materials Engineering, Indian Institute of Technology, Kharagpur 721302, India
*Corresponding author e–mail: aloke@iisc.ac.in

**Abstract**
The interdiffusion coefficients are estimated either following the Wagner's method expressed with respect to the composition (mol or atomic fraction) normalized variable after considering the molar volume variation or the den Broeder's method expressed with respect to the concentration (composition divided by the molar volume) normalized variable. On the other hand, the relations for estimation of the intrinsic diffusion coefficients of components as established by van Loo and integrated diffusion coefficients in a phase with narrow homogeneity range as established by Wagner are currently available with respect to the composition normalized variable only. In this study, we have first derived the relation proposed by den Broeder following the line of treatment proposed by Wagner. Further, the relations for estimation of the intrinsic diffusion coefficients of the components and integrated interdiffusion coefficient are established with respect to the concentration normalized variable, which were not available earlier. The veracity of these methods is examined based on the estimation of data in Ni–Pd, Ni–Al and Cu–Sn systems. Our analysis indicates that both the approaches are logically correct and there is small difference in the estimated data in these systems although a higher difference could be found in other systems. The integrated interdiffusion coefficients with respect to the concentration (or concentration normalized variable) can only be estimated considering the ideal molar volume variation. This might be drawback in certain practical systems.

**Keywords**:
Interdiffusion; Bulk diffusion; Diffusion coefficient estimation.

## 1.     Introduction

Diffusion couple technique is a tool to study diffusion in inhomogeneous materials by coupling dissimilar materials at the temperature of interest [1]. As an added advantage, one can mimic the heterogeneous material systems in application for understanding the phase transformations and the growth of product phases by diffusion–controlled process, which control the various physico–mechanical properties and reliability of the structure [1]. This is even emerged as a research tool to screen a very wide range of compositions optimizing physical and mechanical properties for the development of a new material from only very few samples, which otherwise would need a large volume of samples and unusually high man–time [2].

The two major developments to establish this method as an efficient research tool for diffusion studies can be stated as: (i) The relation developed by Matano [3] for the estimation of composition dependent interdiffusion coefficients. It was developed by simplifying the partial differential equation of Fick's second law [4] to ordinary differential equation utilizing the Boltzmann parameter [5]. This is known as the Matano–Boltzmann analysis. (ii) The Darken–Manning relation [6, 7] developed based on the Kirkendall effect [8] for the estimation of the intrinsic diffusion coefficients (influenced by thermodynamic driving force) and tracer diffusion coefficients (indicating the self–diffusion coefficients) of components [1].

However, the use of Matano–Boltzmann method for the estimation of the interdiffusion coefficients $\widetilde{D}(C_B^*)$ as a function of concentration ($C_B$) introduces error in calculations in most of the practical systems. This relation is expressed as

$$\widetilde{D}(C_B^*) = -\frac{1}{2t\left(\frac{dC_B^*}{dx}\right)}\left[x^*(C_B^* - C_B^-) - \int_{x-\infty}^{x^*}(C_B - C_B^-)\,dx\right] \quad (1)$$

where $t$ is the annealing time and $x^* = (x^* - x_o)$ since the location parameter is measured with respect to $x_o$, i.e., the location of Matano (or initial contact) plane. The asterisk (*) represents the location of interest. Therefore, one of the very important pre–requisites for the use of Matano-Boltzmann analysis is the need to locate the Matano plane. This can be followed only when the molar volume varies ideally with composition or if we consider it as constant. However, it does not fulfill in most of the practical systems and hence, it is almost impossible to locate $x_o$ exactly. As explained mathematically in Ref. [9], it gives different values of $x_o$ when estimated using different components and the difference between them is exactly the same as expansion (for the positive deviation of molar volume) or shrinkage (for the negative deviation of molar volume) of the diffusion couple in a binary system.

To circumvent this problem, mainly two relations are established independently:
(i) The relation developed by Wagner [10] following an analytical approach based on simple algebraic equations, which is expressed as

$$\widetilde{D}(Y_{N_B}^*) = \frac{V_m^*}{2t\left(\frac{dY_{N_B}^*}{dx}\right)}\left[(1 - Y_{N_B}^*)\int_{x-\infty}^{x^*}\frac{Y_{N_B}}{V_m}dx + Y_{N_B}^*\int_{x^*}^{x+\infty}\frac{(1-Y_{N_B})}{V_m}dx\right] \quad (2)$$

where $Y_{N_B} = \frac{N_B - N_B^-}{N_B^+ - N_B^-}$ is the composition normalized variable. $N_B$ is the composition in mol (or atomic) fraction of component B. $V_m$ is the molar volume. "–" and "+" represents the un–reacted left– and right–hand side of the diffusion couple.

(ii) The relation developed by den Broeder [11] by extending the Matano–Boltzmann analysis following a graphical approach, which is expressed as

$$\widetilde{D}(Y_{C_B}^*) = \frac{1}{2t\left(\frac{dY_{C_B}^*}{dx}\right)}\left[(1 - Y_{C_B}^*)\int_{x-\infty}^{x^*}Y_{C_B}\,dx + Y_{C_B}^*\int_{x^*}^{x+\infty}(1 - Y_{C_B})\,dx\right] \quad (3)$$

where $Y_{C_B} = \frac{C_B - C_B^-}{C_B^+ - C_B^-}$ is the concentration normalized variable. $C_B\left(=\frac{N_B}{V_m}\right)$ is the concentration of component B. The main advantage of using any of the above two relations can be understood immediately that there is no need to locate the Matano plane, and hence it can also consider the actual variation of molar volume with composition.

Out of all the methods, the Wagner's method [10] draws a special attention, since in the same manuscript, the author established the concept of the integrated interdiffusion coefficient ($\widetilde{D}_{int}$) for the estimation of the diffusion coefficients in line compounds or the phases with narrow homogeneity range in which concentration gradient cannot be measured. Immediately after that, van Loo [12, 13] proposed the relations for intrinsic $D_i$ (or tracer $D_i^*$) diffusion coefficients of components, in which the Matano plane is not necessary to locate. Much later, Paul [9] derived these relations by extending the Wagner's approach. Both of these relations are derived with the composition normalized variable $Y_{N_B}$.

To summarize, the relations for the estimation of interdiffusion and integrated diffusion coefficients (derived by Wagner [10]), and intrinsic diffusion coefficients (derived by van Loo [13] and Paul [9]) are expressed with respect to composition (mol or atomic fraction) normalized variable $Y_{N_B}$ although the molar volume term to consider the change in total volume of the sample is included correctly during the derivation of these relations (for example, see Equation 2). On the other hand, den Broeder's relation [11] for the interdiffusion coefficient is derived based on concentration (composition divided by molar volume) normalized variable $Y_{C_B}$ in which the molar volume term is automatically included, see Equation 3. The relations for the estimation of other diffusion parameters (integrated and intrinsic diffusion coefficients) with respect to the variable $Y_{C_B}$ are not available. For a constant molar volume, it is easy to visualize from Equations 2 and 3 that both the relations of the interdiffusion coefficients lead to the same equation and therefore will give the same value.

These two methods (den Broeder and Wagner) are compared based on the estimated data only since these are derived completely differently (den Broeder: graphical and Wagner: algebraic formulations). Therefore, with the aim of examining the veracity of these two approaches, we do the following:
(i) For the sake of efficient comparison, we follow the line of treatment proposed by Wagner to check if we can arrive at the den Broeder's relation following Wagner's line of treatment.
(ii) This will then help to extend it to derive the relations for the estimation of the intrinsic diffusion coefficients of components and the integrated interdiffusion coefficient (for the phases with narrow homogeneity range) with respect to $Y_{C_B}$ which are not available at present.
(iii) Following, we consider the experimental results in Ni–Pd (a system with solid solution), Ni–Al (in β–NiAl, a phase with the wide homogeneity range of composition) and Cu–Sn (a system with the narrow homogeneity range phases) to discuss efficiencies/limitations of the approaches.

## 2. Interdiffusion and intrinsic diffusion coefficients with respect to $Y_{C_B}$

The derivation of relations for the interdiffusion coefficients by Wagner [10] and the intrinsic diffusion coefficients by Paul [9] after extending the same line of treatment with respect to composition normalized variable $Y_{N_B}$ can be found in the respective references as mentioned or in the text book as mentioned in Ref. [1]. In this section, we follow the Wagner's line of treatment to find if we can arrive at the den Broeder's relation with respect to $Y_{C_B}$. Then we extend it further to derive the relations for the intrinsic and tracer diffusion coefficients. These will then allow us to compare the data of a particular diffusion parameter when estimated following different relations utilizing $Y_{N_B}$ and $Y_{C_B}$. It should be noted here that the estimation of the tracer diffusion coefficients following the diffusion couple technique is considered indirect but reliable [14],[15],[16],[17],[18]. These are important to correlate the diffusion data with defects assisting the diffusion process in the absence of thermodynamic driving forces.

## 2.1 Derivation of the Interdiffusion Coefficient with respect to $Y_{C_B}$

Interdiffusion coefficients are related to the interdiffusion fluxes following the Fick's first law with respect to component B as [4]

$$\tilde{J}_B = -\tilde{D}\frac{\partial C_B}{\partial x} \tag{4}$$

From the standard thermodynamic relation $C_A\bar{V}_A + C_B\bar{V}_B = 1$ [1], we can write

$$\tilde{D}\frac{\partial C_B}{\partial x} = -\tilde{J}_B = -(C_A\bar{V}_A + C_B\bar{V}_B)\tilde{J}_B \tag{5}$$

where $\bar{V}_i$ are the partial molar volumes of components A and B.

Using another standard thermodynamic equation $\bar{V}_A dC_A + \bar{V}_B dC_B = 0$ [1], we can relate the interdiffusion fluxes with respect to components A and B as

$$\tilde{J}_B = -\tilde{D}\frac{\partial C_B}{\partial x} = \frac{\bar{V}_A}{\bar{V}_B}\tilde{D}\frac{\partial C_A}{\partial x} = -\frac{\bar{V}_A}{\bar{V}_B}\tilde{J}_A$$

$$\bar{V}_B\tilde{J}_B = -\bar{V}_A\tilde{J}_A \tag{6}$$

Note here that the interdiffusion fluxes and the concentration gradients are different at one particular composition (with respect to a particular location in a diffusion couple) in a system with non–ideal molar volume variation. For a constant molar volume $\bar{V}_A = \bar{V}_B = V_m$, these are equal but with opposite sign [19]. On the other hand, the interdiffusion coefficient is the material constant and one will find the same value irrespective of any component considered for the estimation of the data.

Combining Equations (5) and (6), we can write

$$\tilde{D} = \frac{-\tilde{J}_B}{\left(\frac{\partial C_B}{\partial x}\right)} = \frac{-(C_A\bar{V}_A + C_B\bar{V}_B)\tilde{J}_B}{\left(\frac{\partial C_B}{\partial x}\right)}$$

$$\tilde{D} = \frac{\bar{V}_A(C_B\tilde{J}_A - C_A\tilde{J}_B)}{\left(\frac{\partial C_B}{\partial x}\right)}$$

$$\tilde{J}_B = -\bar{V}_A(C_B\tilde{J}_A - C_A\tilde{J}_B) \tag{7}$$

Following Boltzmann [5], compositions in an interdiffusion zone can be related to its position and annealing time by an auxiliary variable as

$$\lambda = \lambda(C_B) = \frac{x-x_o}{\sqrt{t}} = \frac{x}{\sqrt{t}} \tag{8}$$

where $x_o = 0$ is the location of the initial contact plane (Matano plane).

After differentiating Boltzmann parameter in Equation (8) with respect to $t$ and then utilizing the same relation again, we get

$$\frac{d\lambda}{dt} = -\frac{1}{2}\frac{x}{t^{3/2}} = -\frac{\lambda}{2t}$$

$$\frac{-1}{dt} = \frac{\lambda}{2td\lambda} \tag{9}$$

The concentration normalized variable introduced by den Broeder [11] is expressed as

$$Y_{C_B} = \frac{C_B - C_B^-}{C_B^+ - C_B^-} \tag{10}$$

where $C_B^-$ and $C_B^+$ are the concentration of B at the un–affected left– and right–hand side of the diffusion couple.

It can be rearranged to

$$C_B = C_B^+ Y_{C_B} + C_B^- (1 - Y_{C_B}) \tag{11a}$$

Using standard thermodynamic relation $C_A \bar{V}_A + C_B \bar{V}_B = 1$, Equation (11a) can be written as

$$\frac{1 - C_A \bar{V}_A}{\bar{V}_B} = C_B^+ Y_{C_B} + C_B^- (1 - Y_{C_B})$$

$$1 - C_A \bar{V}_A = \bar{V}_B C_B^+ Y_{C_B} + \bar{V}_B C_B^- (1 - Y_{C_B})$$

$$C_A \bar{V}_A = 1 - \bar{V}_B C_B^+ Y_{C_B} - \bar{V}_B C_B^- (1 - Y_{C_B})$$

$$C_A \bar{V}_A = (1 - Y_{C_B}) + Y_{C_B} - \bar{V}_B C_B^+ Y_{C_B} - \bar{V}_B C_B^- (1 - Y_{C_B})$$

$$C_A = \frac{(1 - \bar{V}_B C_B^+) Y_{C_B} + (1 - \bar{V}_B C_B^-)(1 - Y_{C_B})}{\bar{V}_A} \tag{11b}$$

From Fick's second law [4], we know that $\frac{\partial C_i}{\partial t} = \frac{\partial}{\partial x}\left(\tilde{D} \frac{\partial C_i}{\partial x}\right) = -\frac{\partial \tilde{J}_i}{\partial x}$. Therefore, with respect to components A and B and with the help of Equation (9), we can write

$$\frac{\partial \tilde{J}_B}{\partial x} = -\frac{\partial C_B}{\partial t} = \frac{\lambda}{2t} \frac{d(C_B)}{d\lambda} \tag{12a}$$

$$\frac{\partial \tilde{J}_A}{\partial x} = -\frac{\partial C_A}{\partial t} = \frac{\lambda}{2t} \frac{d(C_A)}{d\lambda} \tag{12b}$$

Note here that in Equations (11a) and (11b), the concentrations of component B and A, *i.e.*, $C_B$ and $C_A$ are expressed in terms of the concentration normalized variable ($Y_{C_B}$). So, next we aim to rewrite Fick's second law, *i.e.*, Equations (12a) and (12b) with respect to $Y_{C_B}$.

Replacing Equation (11a) in (12a) and Equation (11b) in (12b), we get

$$\frac{\partial \tilde{J}_B}{\partial x} = \frac{\lambda}{2t}\left[C_B^+ \frac{dY_{C_B}}{d\lambda} + C_B^- \frac{d(1 - Y_{C_B})}{d\lambda}\right] \tag{13a}$$

$$\frac{\partial \tilde{J}_A}{\partial x} = \frac{\lambda}{2t}\left[\left(\frac{1 - \bar{V}_B C_B^+}{\bar{V}_A}\right)\frac{dY_{C_B}}{d\lambda} + \left(\frac{1 - \bar{V}_B C_B^-}{\bar{V}_A}\right)\frac{d(1 - Y_{C_B})}{d\lambda}\right] \tag{13b}$$

Now, we aim to write the above equations with respect to $Y_{C_B}$ and $(1 - Y_{C_B})$ separately.

Operating $[C_B^- \times \text{Eq.}(13b)] - \left[\left(\frac{1 - \bar{V}_B C_B^-}{\bar{V}_A}\right) \times \text{Eq.}(13a)\right]$ leads to

$$C_B^- \frac{\partial \tilde{J}_A}{\partial x} - \left(\frac{1 - \bar{V}_B C_B^-}{\bar{V}_A}\right)\frac{\partial \tilde{J}_B}{\partial x} = \frac{\lambda}{2t}\left(\frac{C_B^- - C_B^+}{\bar{V}_A}\right)\frac{dY_{C_B}}{d\lambda} \tag{14a}$$

Operating $[C_B^+ \times \text{Eq.}(13b)] - \left[\left(\frac{1 - \bar{V}_B C_B^+}{\bar{V}_A}\right) \times \text{Eq.}(13a)\right]$ leads to

$$C_B^+ \frac{\partial \tilde{J}_A}{\partial x} - \left(\frac{1 - \bar{V}_B C_B^+}{\bar{V}_A}\right)\frac{\partial \tilde{J}_B}{\partial x} = \frac{\lambda}{2t}\left(\frac{C_B^+ - C_B^-}{\bar{V}_A}\right)\frac{d(1 - Y_{C_B})}{d\lambda} \tag{14b}$$

After differentiating Boltzmann parameter in Equation (8) with respect to *x*, we get

$$d\lambda = \frac{dx}{\sqrt{t}} \tag{15}$$

Multiplying left–hand side by $\frac{dx}{\sqrt{t}}$ and right–hand side by $d\lambda$ of the Equation (14a) and (14b), respectively, we get

$$\frac{\bar{V}_A C_B^- d\tilde{J}_A - (1 - \bar{V}_B C_B^-) d\tilde{J}_B}{\sqrt{t}} = \left(\frac{C_B^- - C_B^+}{2t}\right)\lambda \, d(Y_{C_B}) \tag{16a}$$

$$\frac{\bar{V}_A C_B^+ d\tilde{J}_A - (1 - \bar{V}_B C_B^+) d\tilde{J}_B}{\sqrt{t}} = \left(\frac{C_B^+ - C_B^-}{2t}\right)\lambda \, d(1 - Y_{C_B}) \tag{16b}$$

Equation (16a) is integrated for a fixed annealing time *t* from un–affected left–hand side of the diffusion couple, *i.e.*, $\lambda = \lambda^{-\infty}$ (corresponds to $x = x^{-\infty}$) to the location of interest $\lambda =$ $\lambda^*$ (corresponds to $x = x^*$) for estimation of the diffusion coefficient. Following, we rearrange, with respect to integration by parts $[\int u dv = uv - \int (v du)]$.

$$\frac{1}{\sqrt{t}}\left[\bar{V}_A C_B^- \int_0^{\tilde{J}_A^*} d\tilde{J}_A - (1 - \bar{V}_B C_B^-)\int_0^{\tilde{J}_B^*} d\tilde{J}_B\right] = \left(\frac{C_B^- - C_B^+}{2t}\right)\int_{\lambda^{-\infty}}^{\lambda^*} \lambda\, d(Y_{C_B})$$

$$\frac{(\bar{V}_A^* C_B^-)\tilde{J}_A^* - (1 - \bar{V}_B^* C_B^-)\tilde{J}_B^*}{\sqrt{t}} = \left(\frac{C_B^- - C_B^+}{2t}\right)\left[\lambda^* Y_{C_B}^* - \int_{\lambda^{-\infty}}^{\lambda^*} Y_{C_B}\, d\lambda\right] \tag{17a}$$

Similarly, Equation (16b) is integrated from the location of interest $\lambda = \lambda^*$ to the un-affected right-hand side of the diffusion couple, i.e., $\lambda = \lambda^{+\infty}$ (corresponds to $x = x^{+\infty}$).

$$\frac{1}{\sqrt{t}}\left[\bar{V}_A C_B^+ \int_{\tilde{J}_A^*}^0 d\tilde{J}_A - (1 - \bar{V}_B C_B^+)\int_{\tilde{J}_B^*}^0 d\tilde{J}_B\right] = \left(\frac{C_B^+ - C_B^-}{2t}\right)\int_{\lambda^*}^{\lambda^{+\infty}} \lambda\, d(1 - Y_{C_B})$$

$$\frac{-(\bar{V}_A^* C_B^+)\tilde{J}_A^* + (1 - \bar{V}_B^* C_B^+)\tilde{J}_B^*}{\sqrt{t}} = \left(\frac{C_B^+ - C_B^-}{2t}\right)\left[-\lambda^*(1 - Y_{C_B}^*) - \int_{\lambda^*}^{\lambda^{+\infty}}(1 - Y_{C_B})\, d\lambda\right] \tag{17b}$$

Note here that the interdiffusion fluxes $\tilde{J}_i$ is equal to zero at the un-affected parts of the diffusion couple, $x = x^{-\infty}$ and $x = x^{+\infty}$, while $\tilde{J}_i^*$ is the fixed value (for certain annealing time $t$) at the location of interest $x = x^*$ in the above Equations (17). Next, we aim to rewrite the above equations with respect to interdiffusion fluxes $\tilde{J}_i$ of both components to get an expression for the interdiffusion coefficient $\tilde{D}$.

Operating $\left[Y_{C_B}^* \times \text{Eq.}(17\text{b})\right] - \left[(1 - Y_{C_B}^*) \times \text{Eq.}(17\text{a})\right]$ leads to

$$\frac{\bar{V}_A^*(C_B^* \tilde{J}_A^* - C_A^* \tilde{J}_B^*)}{\sqrt{t}} = \left(\frac{C_B^+ - C_B^-}{2t}\right)\left[(1 - Y_{C_B}^*)\int_{\lambda^{-\infty}}^{\lambda^*} Y_{C_B}\, d\lambda + Y_{C_B}^* \int_{\lambda^*}^{\lambda^{+\infty}}(1 - Y_{C_B})\, d\lambda\right] \tag{18}$$

Numerator on the left-hand side can be derived, by using $C_B^* = C_B^+ Y_{C_B}^* + C_B^-(1 - Y_{C_B}^*)$ following Equation (11a) and standard thermodynamic relation $\bar{V}_A^* C_A^* = 1 - \bar{V}_B^* C_B^*$, following the steps:

$Y_{C_B}^*\{-(\bar{V}_A^* C_B^+)\tilde{J}_A^* + (1 - \bar{V}_B^* C_B^+)\tilde{J}_B^*\} - (1 - Y_{C_B}^*)\{(\bar{V}_A^* C_B^-)\tilde{J}_A^* - (1 - \bar{V}_B^* C_B^-)\tilde{J}_B^*\}$

$= \{-\bar{V}_A^* C_B^+ Y_{C_B}^* - \bar{V}_A^* C_B^-(1 - Y_{C_B}^*)\}\tilde{J}_A^* + \{(1 - \bar{V}_B^* C_B^+)Y_{C_B}^* + (1 - Y_{C_B}^*)(1 - \bar{V}_B^* C_B^-)\}\tilde{J}_B^*$

$= -\bar{V}_A^*\{C_B^+ Y_{C_B}^* + C_B^-(1 - Y_{C_B}^*)\}\tilde{J}_A^* + \{Y_{C_B}^* - \bar{V}_B^* C_B^+ Y_{C_B}^* + 1 - \bar{V}_B^* C_B^- - Y_{C_B}^* + \bar{V}_B^* C_B^- Y_{C_B}^*\}\tilde{J}_B^*$

$= -\bar{V}_A^*\{C_B^+ Y_{C_B}^* + C_B^-(1 - Y_{C_B}^*)\}\tilde{J}_A^* + [1 - \bar{V}_B^*\{C_B^+ Y_{C_B}^* + C_B^-(1 - Y_{C_B}^*)\}]\tilde{J}_B^*$

$= -\bar{V}_A^*(C_B^* \tilde{J}_A^* - C_A^* \tilde{J}_B^*).$

Utilizing $d\lambda = \frac{dx}{\sqrt{t}}$ from Equation (15), we get

$$\bar{V}_A^*(C_B^* \tilde{J}_A^* - C_A^* \tilde{J}_B^*) = \left(\frac{C_B^+ - C_B^-}{2t}\right)\left[(1 - Y_{C_B}^*)\int_{x^{-\infty}}^{x^*} Y_{C_B}\, dx + Y_{C_B}^* \int_{x^*}^{x^{+\infty}}(1 - Y_{C_B})\, dx\right] \tag{19}$$

For $C_B = C_B^*$, from Equation (7) we know that $\tilde{J}_B^* = -\bar{V}_A^*(C_B^* \tilde{J}_A^* - C_A^* \tilde{J}_B^*)$ and hence the interdiffusion flux with respect to component B can be expressed as

$$\tilde{J}_B^* = \tilde{J}_B(C_B^*) = -\left(\frac{C_B^+ - C_B^-}{2t}\right)\left[(1 - Y_{C_B}^*)\int_{x^{-\infty}}^{x^*} Y_{C_B}\, dx + Y_{C_B}^* \int_{x^*}^{x^{+\infty}}(1 - Y_{C_B})\, dx\right] \tag{20a}$$

Similarly, one can derive the interdiffusion flux with respect to component A as

$$\tilde{J}_A^* = \tilde{J}_A(C_A^*) = \left(\frac{C_A^- - C_A^+}{2t}\right)\left[Y_{C_A}^* \int_{x^{-\infty}}^{x^*}(1 - Y_{C_A})\, dx + (1 - Y_{C_A}^*)\int_{x^*}^{x^{+\infty}} Y_{C_A}\, dx\right] \tag{20b}$$

where $Y_{C_A} = \frac{C_A - C_A^+}{C_A^- - C_A^+}$. Note opposite sign of interdiffusion fluxes when estimated with respect to component A and B because of opposite direction of diffusion of these components.

From Equation (11a) we know that $C_B = C_B^+ Y_{C_B} + C_B^-(1 - Y_{C_B})$. By differentiating it with respect to $x$, we can write

$$\left(\frac{dC_B}{dx}\right)_{x=x^*} = C_B^+ \frac{dY_{C_B}}{dx} - C_B^- \frac{dY_{C_B}}{dx} = (C_B^+ - C_B^-)\left(\frac{dY_{C_B}}{dx}\right)_{x=x^*} \tag{21}$$

From Equation (7) for $C_B = C_B^*$, we know

$$\widetilde{D}(C_B^*) = \frac{-\widetilde{J}_B^*}{\left(\frac{dC_B}{dx}\right)_{x=x^*}} \tag{22}$$

Substituting for flux [Eq. (20a)] and gradient [Eq. (21)] in Fick's first law [Eq. (22)], we get the expression for the estimation of interdiffusion coefficient as

$$\widetilde{D}(Y_{C_B}^*) = \frac{1}{2t\left(\frac{dY_{C_B}^*}{dx}\right)}\left[(1-Y_{C_B}^*)\int_{x-\infty}^{x^*} Y_{C_B}\, dx + Y_{C_B}^* \int_{x^*}^{x+\infty}(1-Y_{C_B})\, dx\right] \tag{23}$$

den Broeder [11] derived this relation with respect to $Y_{C_B}$ following the graphical approach. It should be noted here that the interdiffusion coefficients ($\widetilde{D}(Y_{C_A}^*)$ and $\widetilde{D}(Y_{C_B}^*)$) estimated with respect to component A and B are the same [19]. In this study, we arrive at the same relation (see Equation 3) following the Wagner's [10] line of treatment, although Wagner derived the relation as expressed in Equation 2 with respect to $Y_{N_B}$. Therefore, in a sense, both the relations are logically correct. Additionally, for a constant molar volume ($V_m = V_m^- = V_m^+$), both the den Broeder and the Wagner relations lead to

$$\widetilde{D}(Y_{N_B}^*) = \frac{1}{2t\left(\frac{dY_{N_B}^*}{dx}\right)}\left[(1-Y_{N_B}^*)\int_{x-\infty}^{x^*} Y_{N_B}\, dx + Y_{N_B}^* \int_{x^*}^{x+\infty}(1-Y_{N_B})\, dx\right] \tag{44}$$

Since, $Y_{C_B} = \frac{C_B - C_B^-}{C_B^+ - C_B^-} = \frac{\frac{N_B}{V_m} - \frac{N_B^-}{V_m}}{\frac{N_B^+}{V_m} - \frac{N_B^-}{V_m}} = \frac{N_B - N_B^-}{N_B^+ - N_B^-} = Y_{N_B}$

## 2.2 Derivation of the intrinsic and tracer diffusion coefficients with respect to $Y_{C_B}$

As already mentioned, the relations for the intrinsic diffusion coefficients are available only with respect to $Y_{N_B}$. Therefore, these relations should be derived with respect $Y_{C_B}$ to examine the differences in the data when estimated following these different approaches, i.e., with respect to $Y_{C_B}$ and $Y_{N_B}$. Previously, Paul [9] derived these relations with respect to $Y_{N_B}$ by extending the Wagner's line of treatment to derive the same relations as developed earlier by van Loo [13] differently. We now, extend the analysis to develop the relations for the intrinsic diffusion coefficients with respect to $Y_{C_B}$.

When the location of interest is the position of the Kirkendall marker plane (K), i.e., $x^* = x^K$, we can write Equations (17a) and (17b), respectively as

$$\frac{(\bar{V}_A^K C_B^-)\bar{J}_A^K - (1-\bar{V}_B^K C_B^-)\bar{J}_B^K}{\sqrt{t}} = \left(\frac{C_B^+ - C_B^-}{2t}\right)\left[-\lambda^K Y_{C_B}^K + \int_{\lambda-\infty}^{\lambda^K} Y_{C_B}\, d\lambda\right] \tag{25a}$$

$$\frac{-(\bar{V}_A^K C_B^+)\bar{J}_A^K + (1-\bar{V}_B^K C_B^+)\bar{J}_B^K}{\sqrt{t}} = \left(\frac{C_B^+ - C_B^-}{2t}\right)\left[-\lambda^K(1-Y_{C_B}^K) - \int_{\lambda^K}^{\lambda+\infty}(1-Y_{C_B})\, d\lambda\right] \tag{25b}$$

Now, we aim to rewrite the above equations with respect to $\tilde{J}_B^K$ and $\tilde{J}_A^K$ such that we can get an expression for intrinsic diffusion coefficient of component B and A, i.e., $D_B$ and $D_A$, respectively, at the Kirkendall maker plane utilizing the Darken's equation [6] relating the interdiffusion flux ($\breve{J}_i$) with the intrinsic flux ($J_i$) of component.

Operating $[\bar{V}_A^K C_B^+ \times \text{Eq. (25a)}] + [\bar{V}_A^K C_B^- \times \text{Eq. (25b)}]$ leads to

$$\frac{-(C_B^+ - C_B^-)\tilde{J}_B^K}{\sqrt{t}} = \left(\frac{C_B^+ - C_B^-}{2t}\right)\left[-\lambda^K\{C_B^+ Y_{C_B}^K + C_B^-(1-Y_{C_B}^K)\} + C_B^+ \int_{\lambda-\infty}^{\lambda^K} Y_{C_B}\, d\lambda - C_B^- \int_{\lambda^K}^{\lambda+\infty}(1-Y_{C_B})\, d\lambda\right]$$

Note that $\bar{V}_A^K$ has been cancelled on both sides, since numerator on the left–hand side is $\{-(\bar{V}_A^K C_B^+)(1 - \bar{V}_B^K C_B^-) + \bar{V}_A^K C_B^-(1 - \bar{V}_B^K C_B^+)\}\tilde{J}_B^K = -\bar{V}_A^K(C_B^+ - C_B^-)\tilde{J}_B^K$

Utilizing $C_B^K = C_B^+ Y_{C_B}^K + C_B^-(1 - Y_{C_B}^K)$ from Equation (11a) and after rearranging, we get

$$\tilde{J}_B^K = \frac{\sqrt{t}}{2t}\left[\lambda^K C_B^K - C_B^+ \int_{\lambda^{-\infty}}^{\lambda^K} Y_{C_B} d\lambda + C_B^- \int_{\lambda^K}^{\lambda^{+\infty}} (1 - Y_{C_B}) d\lambda\right] \quad (26a)$$

Similarly, operating $[(1 - \bar{V}_B^K C_B^+) \times \text{Eq.}(25a)] + [(1 - \bar{V}_B^K C_B^-) \times \text{Eq.}(25b)]$ and utilizing $\bar{V}_A^K C_A^K = (1 - \bar{V}_B^K C_B^+) Y_{C_B}^K + (1 - \bar{V}_B^K C_B^-)(1 - Y_{C_B}^K)$ from Equation (11b), we get

$$\frac{-\bar{V}_A^K(C_B^+ - C_B^-)\tilde{J}_A^K}{\sqrt{t}} = \left(\frac{C_B^+ - C_B^-}{2t}\right)\left[-\lambda^K \bar{V}_A^K C_A^K + (1 - \bar{V}_B^K C_B^+) \int_{\lambda^{-\infty}}^{\lambda^K} Y_{C_B} d\lambda - (1 - \bar{V}_B^K C_B^-) \int_{\lambda^K}^{\lambda^{+\infty}} (1 - Y_{C_B}) d\lambda\right]$$

since numerator on the left–hand side is

$\{\bar{V}_A^K C_B^-(1 - \bar{V}_B^K C_B^+) - \bar{V}_A^K C_B^+(1 - \bar{V}_B^K C_B^-)\}\tilde{J}_A^K = -\bar{V}_A^K(C_B^+ - C_B^-)\tilde{J}_A^K$

Dividing both sides of equation by a factor of $\bar{V}_A^K$ and after rearranging, we get

$$\tilde{J}_A^K = \frac{\sqrt{t}}{2t}\left[\lambda^K C_A^K - \left(\frac{1 - \bar{V}_B^K C_B^+}{\bar{V}_A^K}\right)\int_{\lambda^{-\infty}}^{\lambda^K} Y_{C_B} d\lambda + \left(\frac{1 - \bar{V}_B^K C_B^-}{\bar{V}_A^K}\right)\int_{\lambda^K}^{\lambda^{+\infty}} (1 - Y_{C_B}) d\lambda\right] \quad (26b)$$

From Boltzmann parameter in Equation (8), we know that $\lambda^K = \frac{x^K}{\sqrt{t}}$ or $x^K = \lambda^K \sqrt{t}$.

Therefore, the velocity of the Kirkendall marker plane can be expressed as

$$v^K = \frac{dx^K}{dt} = \frac{d(\lambda^K \sqrt{t})}{dt} = \lambda^K \frac{d(\sqrt{t})}{dt} = \frac{\lambda^K}{2\sqrt{t}} = \frac{\lambda^K \sqrt{t}}{2t}$$

Also, differentiating Boltzmann parameter with respect to $x$, from Equation (15) we know that $\sqrt{t}\, d\lambda = dx$.

Putting $\frac{\lambda^K \sqrt{t}}{2t} = v^K$ and $\sqrt{t}\, d\lambda = dx$ in Equations (26), we get

$$\tilde{J}_B^K = v^K C_B^K - \frac{1}{2t}\left[C_B^+ \int_{x^{-\infty}}^{x^K} Y_{C_B} dx - C_B^- \int_{x^K}^{x^{+\infty}} (1 - Y_{C_B}) dx\right] \quad (27a)$$

$$\tilde{J}_A^K = v^K C_A^K - \frac{1}{2t}\left[\left(\frac{1 - \bar{V}_B^K C_B^+}{\bar{V}_A^K}\right)\int_{x^{-\infty}}^{x^K} Y_{C_B} dx - \left(\frac{1 - \bar{V}_B^K C_B^-}{\bar{V}_A^K}\right)\int_{x^K}^{x^{+\infty}} (1 - Y_{C_B}) dx\right] \quad (27b)$$

Following Darken's Analysis [6], we know that $\tilde{J}_B^K = J_B + v^K C_B^K$ and $\tilde{J}_A^K = J_A + v^K C_A^K$. Therefore, we can get an expression for intrinsic flux of component B and A, *i.e.*, $J_B$ and $J_A$, respectively, as follows:

$J_B = \tilde{J}_B^K - v^K C_B^K$

$$J_B = -\frac{1}{2t}\left[C_B^+ \int_{x^{-\infty}}^{x^K} Y_{C_B} dx - C_B^- \int_{x^K}^{x^{+\infty}} (1 - Y_{C_B}) dx\right] \quad (28a)$$

$J_A = \tilde{J}_A^K - v^K C_A^K$

$$J_A = -\frac{1}{2t}\left[\left(\frac{1 - \bar{V}_B^K C_B^+}{\bar{V}_A^K}\right)\int_{x^{-\infty}}^{x^K} Y_{C_B} dx - \left(\frac{1 - \bar{V}_B^K C_B^-}{\bar{V}_A^K}\right)\int_{x^K}^{x^{+\infty}} (1 - Y_{C_B}) dx\right] \quad (28b)$$

Using Fick's first law [4], we can write $D_B = \frac{-J_B}{\left(\frac{\partial C_B}{\partial x}\right)_{x^K}}$ and $D_A = \frac{-J_A}{\left(\frac{\partial C_A}{\partial x}\right)_{x^K}}$. Therefore, we can write an expression for intrinsic diffusion coefficient of component B and A, *i.e.*, $D_B$ and $D_A$, respectively, as follows:

$$D_B = \frac{1}{2t}\left(\frac{\partial x}{\partial C_B}\right)_K \left[C_B^+ \int_{x^{-\infty}}^{x^K} Y_{C_B} dx - C_B^- \int_{x^K}^{x^{+\infty}} (1 - Y_{C_B}) dx\right] \quad (29a)$$

$$D_A = \frac{1}{2t}\left(\frac{\partial x}{\partial C_A}\right)_K \left[\left(\frac{1 - \bar{V}_B^K C_B^+}{\bar{V}_A^K}\right)\int_{x^{-\infty}}^{x^K} Y_{C_B} dx - \left(\frac{1 - \bar{V}_B^K C_B^-}{\bar{V}_A^K}\right)\int_{x^K}^{x^{+\infty}} (1 - Y_{C_B}) dx\right] \quad (29b)$$

The same relation of $D_A$ with respect to $Y_{C_A}$ can be derived as

$$D_A = \frac{1}{2t}\left(\frac{\partial x}{\partial C_A}\right)_K \left[C_A^- \int_{x^{+\infty}}^{x^K} Y_{C_A} dx - C_A^+ \int_{x^K}^{x^{-\infty}} (1 - Y_{C_A}) dx\right] \quad (29c)$$

Compared to Equation 29b, Equation 29c avoids the need for partial molar volumes and hence the error associated with the estimation of these values, as shown later in Section 2.3.

Using $\bar{V}_A dC_A + \bar{V}_B dC_B = 0$, we get $\frac{\partial C_B}{\partial x} = -\frac{\bar{V}_A}{\bar{V}_B}\frac{\partial C_A}{\partial x} \Rightarrow \frac{\partial x}{\partial C_B} = -\frac{\bar{V}_B}{\bar{V}_A}\frac{\partial x}{\partial C_A}$.

Utilizing $\left(\frac{\partial x}{\partial C_B}\right)_K = -\frac{\bar{V}_B^K}{\bar{V}_A^K}\left(\frac{\partial x}{\partial C_A}\right)_K$ in Equations (29), the ratio of intrinsic diffusivities can be written as

$$\frac{D_B}{D_A} = \frac{\bar{V}_B^K}{\bar{V}_A^K}\left[\frac{C_B^+ \int_{x-\infty}^{x^K} Y_{C_B} dx - C_B^- \int_{x^K}^{x+\infty}(1-Y_{C_B})dx}{-C_A^- \int_{x+\infty}^{x^K} Y_{C_A} dx + C_A^+ \int_{x^K}^{x-\infty}(1-Y_{C_A})dx}\right] \tag{29d}$$

This is derived, extending the den Broeder approach for the first time using $Y_{C_B}$. The similar equations with respect to $Y_{N_B}$ as derived by van Loo [13] and Paul [9] are expressed as

$$D_B = \frac{1}{2t}\left(\frac{\partial x}{\partial C_B}\right)_K\left[N_B^+ \int_{x-\infty}^{x^K} \frac{Y_{N_B}}{V_m} dx - N_B^- \int_{x^K}^{x+\infty}\frac{(1-Y_{N_B})}{V_m}dx\right] \tag{30a}$$

$$D_A = \frac{1}{2t}\left(\frac{\partial x}{\partial C_A}\right)_K\left[N_A^+ \int_{x-\infty}^{x^K} \frac{Y_{N_B}}{V_m} dx - N_A^- \int_{x^K}^{x+\infty}\frac{(1-Y_{N_B})}{V_m}dx\right] \tag{30b}$$

$$\frac{D_B}{D_A} = \frac{\bar{V}_B^K}{\bar{V}_A^K}\left[\frac{N_B^+ \int_{x-\infty}^{x^K}\frac{Y_{N_B}}{V_m}dx - N_B^- \int_{x^K}^{x+\infty}\frac{(1-Y_{N_B})}{V_m}dx}{-N_A^+ \int_{x-\infty}^{x^K}\frac{Y_{N_B}}{V_m}dx + N_A^- \int_{x^K}^{x+\infty}\frac{(1-Y_{N_B})}{V_m}dx}\right] \tag{30c}$$

If a constant molar volume is considered (such that the molar volume and the partial molar volumes at every composition are equal, i.e., $V_m = \bar{V}_A = \bar{V}_B$, both the Equations (29) and (30) will be reduced to the same equation

$$D_B = \frac{1}{2t}\left(\frac{\partial x}{\partial N_B}\right)_K\left[N_B^+ \int_{x-\infty}^{x^K} Y_{N_B} dx - N_B^- \int_{x^K}^{x+\infty}(1-Y_{N_B})dx\right] \tag{31a}$$

$$D_A = \frac{1}{2t}\left(\frac{\partial x}{\partial N_A}\right)_K\left[N_A^+ \int_{x-\infty}^{x^K} Y_{N_B} dx - N_A^- \int_{x^K}^{x+\infty}(1-Y_{N_B})dx\right] \tag{31b}$$

Following Darken–Manning Analysis [6, 7], the intrinsic ($D_i$) and tracer ($D_i^*$) diffusion coefficients are related as

$$D_A = \frac{V_m}{\bar{V}_B}D_A^*\Phi(1 + W_A) \tag{32a}$$

$$D_B = \frac{V_m}{\bar{V}_A}D_B^*\Phi(1 - W_B) \tag{32b}$$

where the terms $W_i = \frac{2N_i(D_A^* - D_B^*)}{M_0(N_A D_A^* + N_B D_B^*)}$ arise from the vacancy–wind effect, a constant $M_0$ depends on the crystal structure. $\Phi = \frac{d\ln a_A}{d\ln N_A} = \frac{d\ln a_B}{d\ln N_B}$ is the thermodynamic factor which (according to the Gibbs–Duhem relation) is same for both the components A and B in a binary system. $a_i$ is the activity of component $i$. Therefore, the tracer diffusion coefficients can be estimated from the known thermodynamic parameters following Equations 29 or 30 and 32.

## 2.3 Comparison of the interdiffusion and intrinsic diffusion coefficients estimated following the relations established with respect to $Y_{C_B}$ and $Y_{N_B}$

We compare the estimated values based on the estimation of diffusion coefficients in the Ni–Pd system [20]. The interdiffusion zone developed after annealing Ni and Pd at 1100 °C for 196 hrs is shown in Figure 1a. The location of the Kirkendall marker plane is identified by the ThO$_2$ particles, at 40.3 at% Ni. The composition profile developed in the interdiffusion zone is shown in Figure 1b. This is measured in a direction perpendicular to the Kirkendall marker plane following the diffusion direction of the components. The variation of molar volume used for the estimation of diffusion coefficients is shown in Figure 1c. The partial molar volumes of the components at the composition of the Kirkendall marker plane ($N_{Ni}^K = 0.403$) are shown. Since this diffusion couple is prepared with pure components as the end–members, the composition normalized variables are the same as composition of the respective components, as shown in Figure 1b. The concentration normalized variables for component A and B are shown in Figure 1d. The estimated data are shown in Figure 2. To compare the data, the different parts of the den Broeder's and the Wagner's relations are plotted. Gradients of concentration normalized variable $\left(\frac{dY_{C_B}^*}{dx}\right)$ and composition normalized variable $\left(\frac{dY_{N_B}^*}{dx}\right)$ are shown in Figure 2a. The bracketed terms $\left[(1 - Y_{C_B}^*)\int_{x-\infty}^{x^*} Y_{C_B} dx + Y_{C_B}^* \int_{x^*}^{x+\infty}(1 - Y_{C_B}) dx\right] = 2t \times \widetilde{D}(Y_{C_B}^*) \times \left(\frac{dY_{C_B}^*}{dx}\right)$ and $V_m^*\left[(1 - Y_{N_B}^*)\int_{x-\infty}^{x^*}\frac{Y_{N_B}}{V_m} dx + Y_{N_B}^* \int_{x^*}^{x+\infty}\frac{(1-Y_{N_B})}{V_m} dx\right] = 2t \times \widetilde{D}(Y_{N_B}^*) \times \left(\frac{dY_{N_B}^*}{dx}\right)$ are shown in Figure 2b. Following, $\widetilde{D}(Y_{C_B}^*)$ and $\widetilde{D}(Y_{N_B}^*)$ are shown in Figure 2c. As expected based on the definition of terms $Y_{C_B}$ and $Y_{N_B}$, although there is difference in the slope and the bracket terms; however, a very minor difference in the estimated diffusion coefficients with respect to $Y_{C_B}$ and $Y_{N_B}$ is evident.

Following, the intrinsic diffusion coefficients of components following den Broeder and Wagner methods are estimated. Since pure end–members are used, considering the composition profile in Figure 1b, we can write $N_B^-(=N_{Ni}^-) = 0$, $N_A^+(=N_{Pd}^+) = 0$, $N_B^+(=N_{Ni}^+) = 1$, and $N_A^-(=N_{Pd}^-) = 1$. Therefore, we have $C_A^-(=C_{Pd}^-) = \frac{N_{Pd}^-}{V_m^-} = \frac{1}{V_{Pd}}$, $C_B^-(=C_{Ni}^-) = \frac{N_{Ni}^-}{V_m^-} = \frac{0}{V_{Pd}} = 0$, $C_A^+(=C_{Pd}^+) = \frac{N_{Pd}^+}{V_m^+} = \frac{0}{V_{Ni}} = 0$ and $C_B^+(=C_{Ni}^+) = \frac{N_{Ni}^+}{V_m^+} = \frac{1}{V_{Ni}}$. Therefore, we can simplify the Equation 29a, b and c in the case of Ni–Pd diffusion couple as

$D_B(=D_{Ni}) = \frac{1}{2t}\left(\frac{\partial x}{\partial C_B}\right)_K \left[C_B^+ \int_{x-\infty}^{x^K} Y_{C_B} dx\right] = \frac{1}{2t}\left(\frac{\partial x}{\partial C_{Ni}}\right)_K \left[\frac{1}{V_{Ni}} \int_{x-\infty}^{x^K} Y_{C_{Ni}} dx\right] = 2.6 \times 10^{-14}$ m$^2$/s,

$D_A(=D_{Pd}) = \frac{1}{2t}\left(\frac{\partial x}{\partial C_A}\right)_K \left[\left(\frac{1 - \overline{V}_B^K C_B^+}{\overline{V}_A^K}\right)\int_{x-\infty}^{x^K} Y_{C_B} dx - \left(\frac{1}{\overline{V}_A^K}\right)\int_{x^K}^{x+\infty}(1 - Y_{C_B}) dx\right] =$

$\frac{1}{2t}\left(\frac{\partial x}{\partial C_{Pd}}\right)_K \left[\left(\frac{1 - \frac{\overline{V}_{Ni}^K}{V_{Ni}}}{\overline{V}_{Pd}^K}\right)\int_{x-\infty}^{x^K} Y_{C_{Ni}} dx - \left(\frac{1}{\overline{V}_{Pd}^K}\right)\int_{x^K}^{x+\infty}(1 - Y_{C_{Ni}}) dx\right] = 5.2 \times 10^{-14}$ m$^2$/s,

$D_A(=D_{Pd}) = \frac{1}{2t}\left(\frac{\partial x}{\partial C_A}\right)_K \left[C_A^- \int_{x+\infty}^{x^K} Y_{C_A} dx\right] = \frac{1}{2t}\left(\frac{\partial x}{\partial C_{Pd}}\right)_K \left[\frac{1}{V_{Pd}} \int_{x+\infty}^{x^K} Y_{C_{Pd}} dx\right] = 4.9 \times 10^{-14}$ m$^2$/s.

The same can be estimated following the Wagner's method modifying the Equations 30a and b for the Ni–Pd diffusion couple at the Kirkendall marker plane as

$D_B(=D_{Ni}) = \frac{1}{2t}\left(\frac{\partial x}{\partial C_B}\right)_K \left[\int_{x-\infty}^{x^K} \frac{Y_{N_B}}{V_m} dx\right] = \frac{1}{2t}\left(\frac{\partial x}{\partial C_{Ni}}\right)_K \left[\int_{x-\infty}^{x^K} \frac{Y_{N_{Ni}}}{V_m} dx\right] = 2.6 \times 10^{-14}$ m$^2$/s,

$D_A(=D_{Pd}) = \frac{1}{2t}\left(\frac{\partial x}{\partial C_A}\right)_K \left[-\int_{x^K}^{x+\infty}\frac{(1-Y_{N_B})}{V_m} dx\right] = \frac{1}{2t}\left(\frac{\partial x}{\partial C_{Pd}}\right)_K \left[-\int_{x^K}^{x+\infty}\frac{(1-Y_{N_{Ni}})}{V_m} dx\right] = 4.9 \times 10^{-14}$ m$^2$/s.

Therefore, there is no difference in the estimated intrinsic diffusion coefficients following den Broeder and Wagner methods. A small difference in values of intrinsic diffusion coefficient of Pd is found following the den Broeder method when estimated with respect to $Y_{C_{Ni}}$ and $Y_{C_{Pd}}$. This must be because of error associated with the calculation of partial molar volumes while estimating the data utilizing $Y_{C_{Ni}}$.

Following, we estimate the interdiffusion coefficients in the β–NiAl phase. The composition profile of a diffusion couple $Ni_{0.46}Al_{0.54}$ / $Ni_{0.575}Al_{0.425}$ after annealing at 1200 °C for 24 hrs is shown in Figure 3a. The molar volume variation in this intermetallic compound is shown in Figure 3b [9]. The estimated interdiffusion coefficients by two methods are shown in Figure 3c. The difference between the data estimated using both the methods in this system is higher compared to the Ni–Pd system.

## 3. Integrated Interdiffusion Coefficient

Wagner, in his seminal contribution [10], introduced the concept of the integrated interdiffusion coefficient in a phase with narrow homogeneity range since the concentration/composition gradient in such a phase cannot be determined. This is expressed with respect to $Y_{N_B}$ or $N_B$ as

$$\widetilde{D}_{int}^{\beta} = V_m^{\beta} \Delta x^{\beta} \left( \frac{N_B^+ - N_B^-}{2t} \right) \left[ \frac{Y_{N_B}^{\beta}\left(1 - Y_{N_B}^{\beta}\right)}{V_m^{\beta}} \Delta x^{\beta} + \left(1 - Y_{N_B}^{\beta}\right) \int_{x^{-\infty}}^{x^{\beta_1}} \frac{Y_{N_B}}{V_m} dx + Y_{N_B}^{\beta} \int_{x^{\beta_2}}^{x^{+\infty}} \frac{(1 - Y_{N_B})}{V_m} dx \right]$$
(33a)

$$\widetilde{D}_{int}^{\beta} = \frac{(N_B^{\beta} - N_B^-)(N_B^+ - N_B^{\beta})}{(N_B^+ - N_B^-)} \frac{(\Delta x^{\beta})^2}{2t}$$
$$+ \frac{V_m^{\beta} \Delta x^{\beta}}{2t} \left[ \frac{(N_B^+ - N_B^{\beta})}{(N_B^+ - N_B^-)} \int_{x^{-\infty}}^{x^{\beta_1}} \frac{(N_B - N_B^-)}{V_m} dx + \frac{(N_B^{\beta} - N_B^-)}{(N_B^+ - N_B^-)} \int_{x^{\beta_2}}^{x^{+\infty}} \frac{(N_B^+ - N_B)}{V_m} dx \right]$$
(33b)

At present, the relation to estimate the same diffusion parameter with respect to $Y_{C_B}$ or $C_B$ is not available. Therefore, as given in the supplementary file, we derived this relation by extending the den Broeder's relation for the interdiffusion coefficient. This is expressed with respect to $Y_{C_B}$ or $C_B$ as

$$\widetilde{D}_{int}^{\beta} = \frac{(V_m^{\beta})^2}{\overline{V}_A^{\beta}} \Delta x^{\beta} \left( \frac{C_B^+ - C_B^-}{2t} \right) \left[ Y_{C_B}^{\beta}\left(1 - Y_{C_B}^{\beta}\right) \Delta x^{\beta} + \left(1 - Y_{C_B}^{\beta}\right) \int_{x^{-\infty}}^{x^{\beta_1}} Y_{C_B} dx + Y_{C_B}^{\beta} \int_{x^{\beta_2}}^{x^{+\infty}} (1 - Y_{C_B}) dx \right]$$
(34a)

$$\widetilde{D}_{int}^{\beta} = \frac{(V_m^{\beta})^2}{\overline{V}_A^{\beta}} \frac{(C_B^{\beta} - C_B^-)(C_B^+ - C_B^{\beta})}{(C_B^+ - C_B^-)} \frac{(\Delta x^{\beta})^2}{2t} + \frac{(V_m^{\beta})^2}{\overline{V}_A^{\beta}} \frac{\Delta x^{\beta}}{2t} \left[ \frac{C_B^+ - C_B^{\beta}}{C_B^+ - C_B^-} \int_{x^{-\infty}}^{x^{\beta_1}} (C_B - C_B^-) dx + \frac{C_B^{\beta} - C_B^-}{C_B^+ - C_B^-} \int_{x^{\beta_2}}^{x^{+\infty}} (C_B^+ - C_B) dx \right]$$
(34b)

It can be seen that an additional term of partial molar volume is present in the relation expressed with respect to $Y_{C_B}$ or $C_B$ as compared to the relation expressed with respect to $Y_{N_B}$ or $N_B$.

Now we compare the efficiencies and difficulties of estimation of the data utilizing the growth of the product phases, as shown in Figure 4a, in the Cu–Sn system. The Cu/Sn diffusion couple was annealed at 200 °C for 81 hrs (*i.e.*, $2t = 2 \times 81 \times 3600$ s) in which two phases Cu$_3$Sn and Cu$_6$Sn$_5$ grows in the interdiffusion zone [16]. The average thicknesses of the phases are estimated as 3.5 μm ($= \Delta x^{Cu_3Sn}$) for Cu$_3$Sn and 13 μm ($= \Delta x^{Cu_6Sn_5}$) for Cu$_6$Sn$_5$. The marker plane, detected by the presence of duplex morphology, in the Cu$_6$Sn$_5$ phase is found at a distance of 7 μm from the Cu$_3$Sn/Cu$_6$Sn$_5$ interface. The actual molar volumes of these phases are estimated as $V_m^{Cu_3Sn} = 8.59 \times 10^{-6}$ and $V_m^{Cu_6Sn_5} = 10.59 \times 10^{-6}$ m$^3$/mol. From the knowledge of the molar volumes of the end–member components $V_m^- = V_m^{Cu} = 7.12 \times 10^{-6}$ and $V_m^+ = V_m^{Sn} = 16.24 \times 10^{-6}$ m$^3$/mol, we can estimate the ideal molar volume of the product phase of interest ($\beta$) following the Vegard's law $V_m^\beta = N_{Cu}^\beta V_m^{Cu} + N_{Sn}^\beta V_m^{Sn}$. This is estimated as $V_m^{Cu_3Sn}(ideal) = 9.4 \times 10^{-6}$ m$^3$/mol for the Cu$_3$Sn phase $\left(N_{Cu}^{Cu_3Sn} = \frac{3}{4}, N_{Sn}^{Cu_3Sn} = \frac{1}{4}\right)$ and $V_m^{Cu_6Sn_5}(ideal) = 11.3 \times 10^{-6}$ m$^3$/mol for the Cu$_6$Sn$_5$ phase $\left(N_{Cu}^{Cu_6Sn_5} = \frac{6}{11}, N_{Sn}^{Cu_6Sn_5} = \frac{5}{11}\right)$. Therefore, as shown in Figure 4b, the negative deviations of the molar volumes are 8.6% for the Cu$_3$Sn phase and 6.3% for the Cu$_6$Sn$_5$ phase.

The detailed estimation procedure following the Wagner method can be found in books as mentioned in Refs. [1, 21]. As explained in detail in the supplementary file, the integrated diffusion coefficients of the phases following this method are estimated as $\widetilde{D}_{int}^{Cu_3Sn} = 1.26 \times 10^{-17}$ $m^2/s$ and $\widetilde{D}_{int}^{Cu_6Sn_5} = 8.49 \times 10^{-17}$ $m^2/s$. As it should be, the same values are estimated considering the components A and B following Equations S10 or S11 in the supplementary file. The ratio of diffusivities in the Cu$_6$Sn$_5$ phase is estimated as $\frac{D_{Sn}^*}{D_{Cu}^*} = 1.30 \pm 0.05$. It should be noted here that a different value of this ratio was reported in Ref. [16], which was an average of data estimated at different locations in different diffusion couples, compared to the data reported in this study estimated based on the micrograph, as shown in Figure 4a.

Compared to the Wagner method (Equations S10 or S11 in the supplementary file), den Broeder method (Equations S7 or S8 in the supplementary file) has an additional complication because of the presence of partial molar volume terms in them. In a compound with narrow homogeneity range, the variation of the lattice parameter with respect to the composition is not known. The variation in such a small composition range might be small; however, the difference between the partial molar volumes could still be very high. To circumvent this problem, there could be two options: (i) consider $V_m^\beta = \overline{V}_A^\beta = \overline{V}_B^\beta$, *i.e.*, a constant molar volume in the phase of interest β or (ii) an ideal variation of the molar volume in the whole A–B system. To discuss the pros and cons of these two assumptions, we extend our analysis based on the estimated data in the Cu–Sn system. Following the first assumption, as listed in column number 2 and 3 of Table 1, we estimate two different values of the data when estimated following the composition/concentration profile of component A and B. The estimation steps can be found in the supplementary file. This comes from the fact that the assumption leads to different (absolute) values of the interdiffusion fluxes, *i.e.*, $\left|\tilde{J}_A^\beta\right| \neq \left|\tilde{J}_B^\beta\right|$, when estimated following the Equations S6a and S6b (see supplementary file) because of this assumption. Moreover, when a constant molar volume is considered, following Equation 6, we should have $\bar{V}_B\tilde{J}_B + \bar{V}_A\tilde{J}_A = \tilde{J}_A^\beta + \tilde{J}_B^\beta = 0$. In fact, the assumptions should be taken such that this relation is fulfilled. Therefore, this is not a valid assumption for the estimation of the integrated diffusion coefficients following the den Broeder method, *i.e.*, relations with respect to concentration normalized variable.

Therefore, the den Broeder method for estimation of the integrated diffusion coefficients can be used considering an ideal variation of the molar volume, as shown by dotted line in Figure 4b. This fulfills the condition $\bar{V}_B\tilde{J}_B + \bar{V}_A\tilde{J}_A = 0$, where the partial molar volumes are equal to the molar volumes of the end–member components. Following, we get a same value of the integrated diffusion coefficient in a particular phase as $\tilde{D}_{int}^{Cu_3Sn} = 1.28\times 10^{-17}\ m^2/s$ and $\tilde{D}_{int}^{Cu_6Sn_5} = 8.53\times 10^{-17}\ m^2/s$. The ratio of diffusivities $\frac{D_{Sn}^*}{D_{Cu}^*}$ is found to be $1.29 \pm 0.05$. It can be seen in Table 1 that there is very small difference in the estimated values following Wagner and den Broeder method. Therefore, one can practically follow any of the methods. However, it is advisable to follow the Wagner method since there is no need of considering the ideal molar volume variation instead of considering the actual molar volume variation, which might play a significant effect in certain systems.

## 4. Conclusion

The relation for the composition dependent interdiffusion coefficient was first proposed by Matano [3] in 1933, which was difficult to follow in most of the practical systems. As a result, there were many efforts to develop a better relation. Balluffi [22], Sauer–Freise [23], Wagner [10] and den Broeder [11] proposed relations, which played influential role in the field of solid–state diffusion. Currently, two approaches are followed with equal importance by different groups. One was proposed by Wagner with respect to composition normalized variable after considering the molar volume variation and another one was proposed by den Broeder with respect to the concentration normalized variable. Although, it is known to produce different values of the interdiffusion coefficient depending on the molar volume variation [24], the choice of a method by a particular research group is rather random. Incidentally both the methods were published in the same year 1969. The manuscript published by Wagner draws special attention since he put forward the concept of the integrated diffusion coefficient for the phases with narrow homogeneity range in which the interdiffusion coefficients cannot be determined because of unknown composition (or concentration) gradient. This relation is therefore naturally derived with respect to the composition normalized variable. Even the relations for the estimation of the intrinsic diffusion coefficients were also derived by van Loo [13] with respect to the composition normalized variable, which was later derived again by Paul [9] extending the Wagner's analysis.

To examine the veracity of the methods with respect to composition and concentration normalized variables, the relation proposed by den Broeder is first derived following the line of treatment followed by Wagner. Following, this is extended to derive the relations for the intrinsic diffusion coefficients and the integrated diffusion coefficients to develop the relations with respect to the concentration normalized variable, which were not available earlier. We have shown further that an additional assumption of the ideal molar volume variation is required for the estimation of the integrated diffusion coefficient with respect to the concentration normalized variable when compared to the relation developed by Wagner with respect to the composition normalized variable, which can be used with actual molar volume variation.

**Acknowledgement**

We acknowledge the financial support from ARDB, India (Project number ARDB/GTMAP/01/2031786/M/1, DT. 11/1/2016).

## References

[1]     A. Paul, T. Laurila, V. Vuorinen, S.V. Divinski, Thermodynamics, Diffusion and the Kirkendall Effect in Solids, 1st ed., Springer, Switzerland, 2014.
[2]     A. Kodentsov, A. Paul, Chapter 6 - Diffusion Couple Technique: A Research Tool in Materials Science, Handbook of Solid State Diffusion, Editors: A. Paul and S. Divinsky, Volume 2: Diffusion Analysis and Material Applications, Elsevier 2017, pp. 207-275.
[3]     C. Matano, On the relation between the diffusion-coefficients and concentrations of solid metals (the nickel-copper system), Japanese Journal of Physics 8 (1933) 109-113.
[4]     A. Fick, Ueber Diffusion, Annalen der Physik 170 (1855) 59-86.
[5]     L. Boltzmann, Zur Integration der Diffusionsgleichung bei variabeln Diffusions-coefficienten, Annalen der Physik 289 (1894) 959-964.
[6]     L.S. Darken, Diffusion, Mobility and Their Interrelation through Free Energy in Binary Metallic Systems, Transactions of the American Institute of Mining and Metallurgical Engineers 175 (1948) 184-201.
[7]     J.R. Manning, Diffusion and the Kirkendall shift in binary alloys, Acta Metallurgica 15 (1967) 817-826.
[8]     A.D. Smigelskas, E.O. Kirkendall, Zinc Diffusion in Alpha-Brass, Transactions of the American Institute of Mining and Metallurgical Engineers 171 (1947) 130-142.
[9]     A. Paul, PhD Thesis, The Kirkendall effect in solid state diffusion, Eindhoven University of Technology, The Netherlands, 2004. Available at alexandria.tue.nl/extra2/200412361.pdf.
[10]    C. Wagner, The evaluation of data obtained with diffusion couples of binary single-phase and multiphase systems, Acta Metallurgica 17 (1969) 99-107.
[11]    F.J.A. den Broeder, A general simplification and improvement of the Matano-Boltzmann method in the determination of the interdiffusion coefficients in binary systems, Scripta Metallurgica 3 (1969) 321-325.
[12]    F.J.J. Van Loo, On the determination of diffusion coefficients in a binary metal system, Acta Metallurgica 18 (1970) 1107-1111.
[13]    F.J.J. van Loo, Multiphase diffusion in binary and ternary solid-state systems, Progress in Solid State Chemistry 20 (1990) 47-99.

[14] V.A. Baheti, S. Roy, R. Ravi, A. Paul, Interdiffusion and the phase boundary compositions in the Co–Ta system, Intermetallics 33 (2013) 87-91.

[15] V.A. Baheti, R. Ravi, A. Paul, Interdiffusion study in the Pd–Pt system, Journal of Materials Science: Materials in Electronics 24 (2013) 2833-2838.

[16] V.A. Baheti, S. Kashyap, P. Kumar, K. Chattopadhyay, A. Paul, Bifurcation of the Kirkendall marker plane and the role of Ni and other impurities on the growth of Kirkendall voids in the Cu–Sn system, Acta Materialia 131 (2017) 260-270.

[17] V.A. Baheti, S. Kashyap, P. Kumar, K. Chattopadhyay, A. Paul, Solid–state diffusion–controlled growth of the phases in the Au–Sn system, Philosophical Magazine 98(1) (2018) 20-36.

[18] A. Paul, C. Ghosh, W.J. Boettinger, Diffusion Parameters and Growth Mechanism of Phases in the Cu–Sn System, Metallurgical and Materials Transactions A 42 (2011) 952-963.

[19] A. Paul, Comments on "Sluggish diffusion in Co–Cr–Fe–Mn–Ni high-entropy alloys" by K.Y. Tsai, M.H. Tsai and J.W. Yeh, Acta Materialia 61 (2013) 4887–4897, Scripta Materialia 135 (2017) 153-157.

[20] M.J.H. van Dal, PhD Thesis, Microstructural stability of the Kirkendall plane, Eindhoven University of Technology, The Netherlands, 2001.

[21] A. Paul, Chapter 3 - Estimation of Diffusion Coefficients in Binary and Pseudo-Binary Bulk Diffusion Couples, in: Handbook of Solid State Diffusion, Editors: A. Paul and S. Divinsky, Volume 1: Diffusion Fundamentals and Techniques, Elsevier 2017, pp. 79-201.

[22] R.W. Balluffi, On the determination of diffusion coefficients in chemical diffusion, Acta Metallurgica 8 (1960) 871-873.

[23] F. Sauer, V. Freise, Diffusion in binären Gemischen mit Volumenänderung, Zeitschrift für Elektrochemie, Berichte der Bunsengesellschaft für physikalische Chemie 66 (1962) 353-362.

[24] S. Santra, A. Paul, Role of the Molar Volume on Estimated Diffusion Coefficients, Metallurgical and Materials Transactions A 46 (2015) 3887-3899.

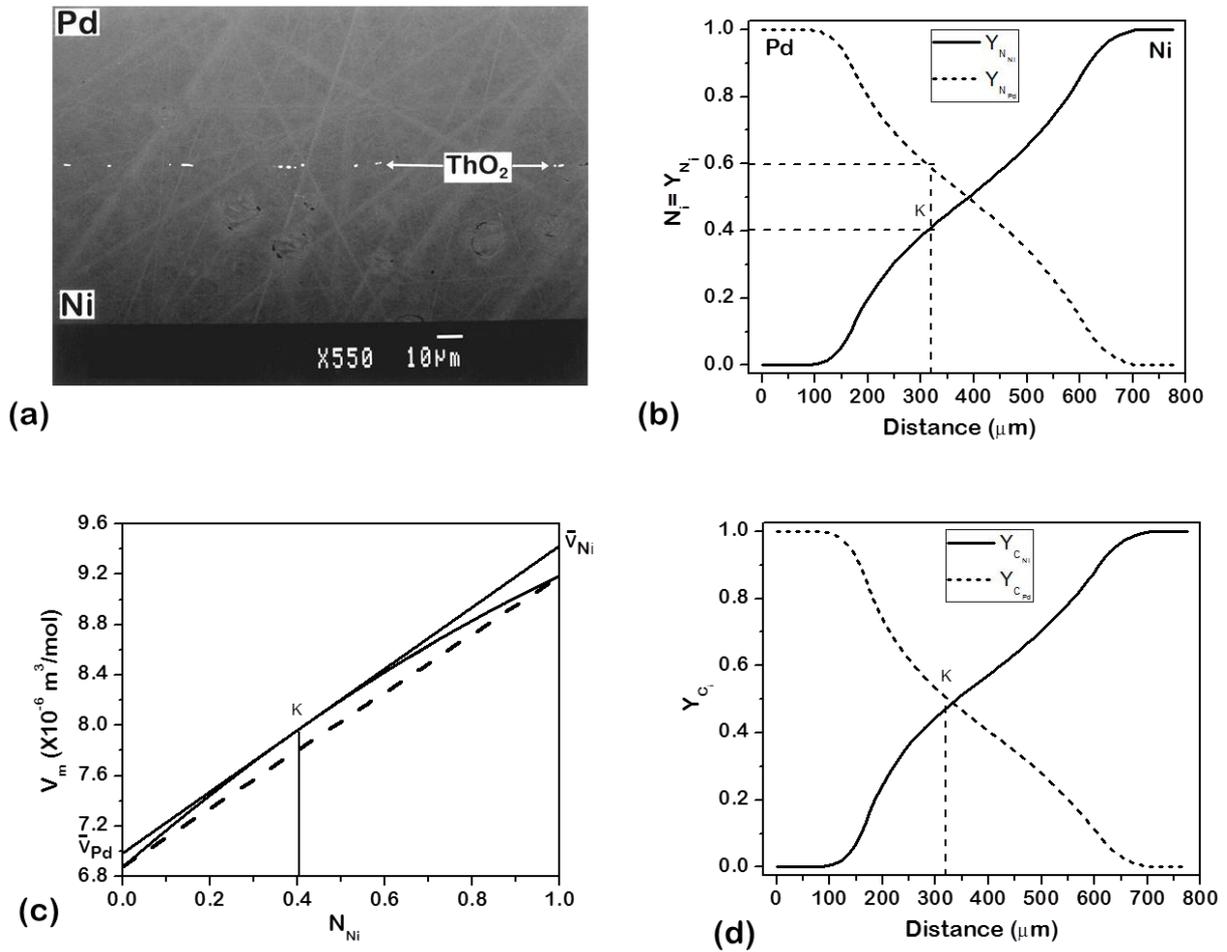

Figure 1: (a) Micrograph of the Ni/Pd diffusion couple annealed at 1100 °C for 196 hrs [20]. ThO$_2$ particles identifies the location of the Kirkendall marker plane, (b) the corresponding composition profile [20] (equal to the composition normalize variable, $Y_{N_B}$) developed in the interdiffusion zone, (c) Molar volume variation in the Ni–Pd solid solution [20], (d) the concentration normalized variables.

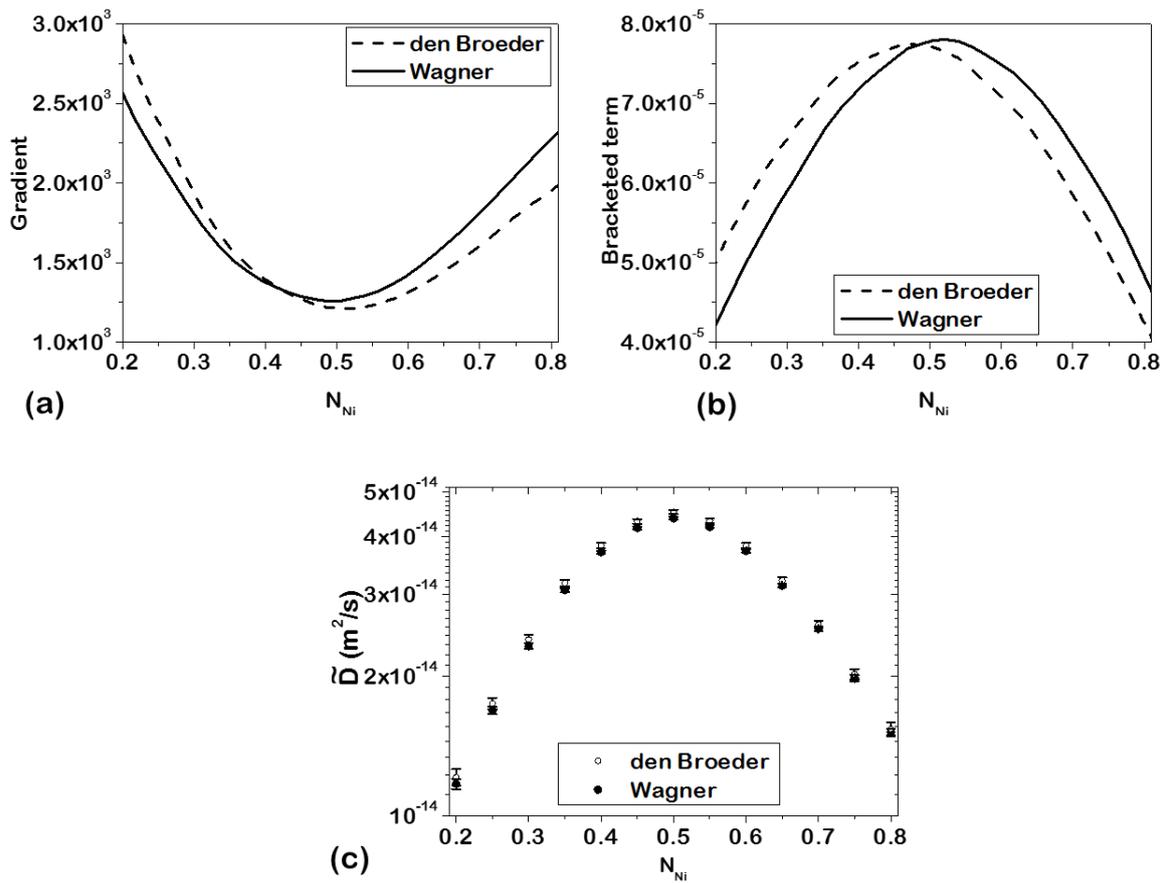

Figure 2: Estimated data in the Ni–Pd system using the Wagner and the den Broeder methods: (a) Gradients $\left(\frac{dY^*_{N_B}}{dx}\right)$ for Wagner relation and $\left(\frac{dY^*_{C_B}}{dx}\right)$ for den Broeder relation (b) bracketed terms of the Wagner and den Broeder relations and (c) the estimated interdiffusion coefficients.

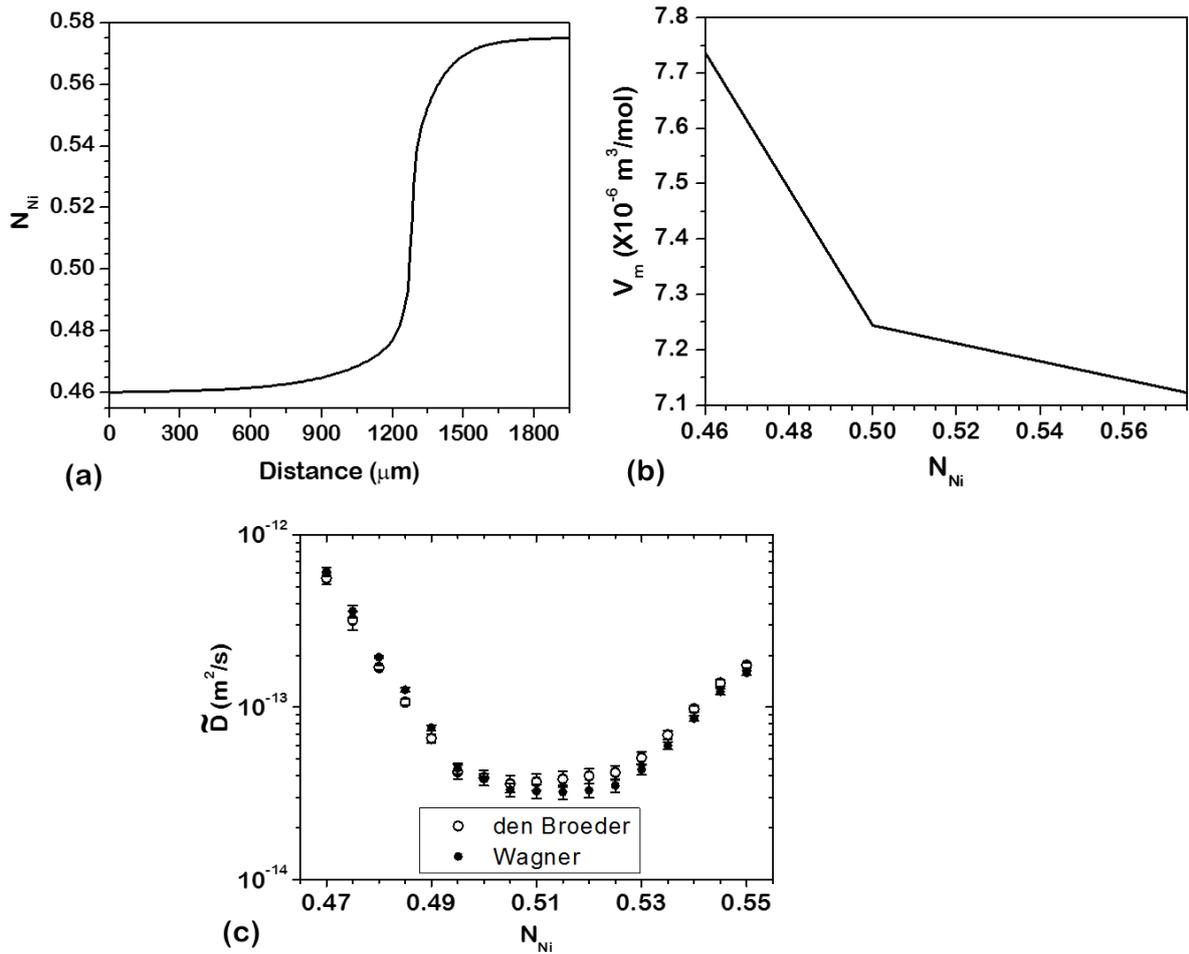

Figure 3: (a) Composition profile of β–NiAl phase grown in $Ni_{0.46}Al_{0.54}$ / $Ni_{0.575}Al_{0.425}$ diffusion couple after annealing at 1200 °C for 24 hrs, (b) the molar volume variation in the β–NiAl phase [9] and (c) the estimated interdiffusion coefficients following the Wagner and the den Broeder approaches.

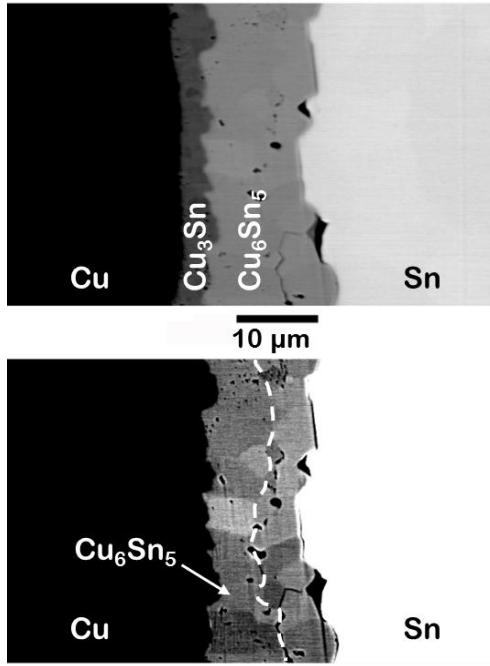 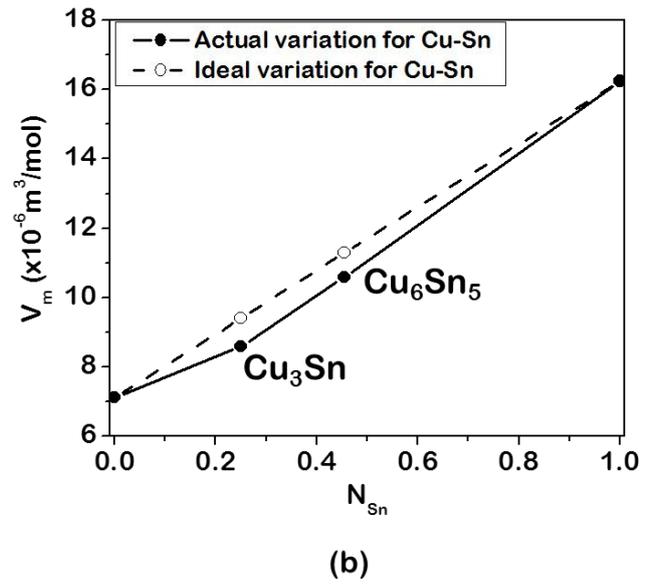

(a)

Figure 4: (a) BSE micrographs showing (top) the growth of $Cu_3Sn$ and $Cu_6Sn_5$ phases in the interdiffusion zone of Cu/Sn diffusion couple annealed at 200 °C for 81 hrs [16], where (bottom) the location of the Kirkendall marker plane as denoted by a dashed line is indicated by duplex morphology inside $Cu_6Sn_5$, and (b) actual as well as ideal variations of molar volume in the Cu–Sn system.

**Table**

| | Following composition normalize variable | Following concentration normalize variable | | |
|---|---|---|---|---|
| | Using $Y_{N_{Sn}}$ or $Y_{N_{Cu}}$<br>Actual $V_m$ | Using $Y_{C_{Sn}}$<br>Actual $V_m$ | Using $Y_{C_{Cu}}$<br>Actual $V_m$ | Using $Y_{C_{Sn}}$<br>Ideal $V_m$ |
| $\widetilde{D}_{int}^{Cu_3Sn}$<br>($\times 10^{-17}$ m²/s) | $1.26 \pm 0.05$ | $\frac{V_m}{\overline{V}_{Cu}} 8.69 \times 10^{-18}$<br>or $(8.69 \pm 0.05) \times 10^{-18}$<br>Considering $\overline{V}_i^\beta = V_m^\beta$ | $\frac{V_m}{\overline{V}_{Sn}} 1.90 \times 10^{-17}$<br>or $(1.90 \pm 0.05) \times 10^{-17}$<br>Considering $\overline{V}_i^\beta = V_m^\beta$ | $1.28 \pm 0.05$ |
| $\widetilde{D}_{int}^{Cu_6Sn_5}$<br>($\times 10^{-17}$ m²/s) | $8.49 \pm 1$ | $\frac{V_m}{\overline{V}_{Cu}} 4.72 \times 10^{-17}$<br>or $(4.72 \pm 1) \times 10^{-17}$<br>Considering $\overline{V}_i^\beta = V_m^\beta$ | $\frac{V_m}{\overline{V}_{Sn}} 1.16 \times 10^{-16}$<br>or $(1.16 \pm 1) \times 10^{-16}$<br>Considering $\overline{V}_i^\beta = V_m^\beta$ | $8.53 \pm 1$ |

Table 1: Diffusion parameters estimated in the Cu₃Sn and Cu₆Sn₅ phases using Cu and Sn profiles following both the Wagner and the den Broeder methods using the actual as well as the ideal variation of molar volumes in the Cu/Sn diffusion couple.

# Supplementary File

## S.1 Derivation of the relation for the Integrated Interdiffusion Coefficient with respect to concentration normalized variable as well as composition normalized variable

The integrated diffusion coefficient $\left(\widetilde{D}_{int}^{\beta}\right)$ in a phase (β) with narrow homogeneity range is defined as the interdiffusion coefficient $(\widetilde{D})$ integrated over the unknown composition range of the phase of interest such that

$$\widetilde{D}_{int}^{\beta} = \int_{N_B^{\beta_1}}^{N_B^{\beta_2}} \widetilde{D} \, dN_B = \widetilde{D} \int_{N_B^{\beta_1}}^{N_B^{\beta_2}} dN_B = \widetilde{D}\left(N_B^{\beta_2} - N_B^{\beta_1}\right)$$

$$\widetilde{D}_{int}^{\beta} = \widetilde{D} \Delta N_B^{\beta} \tag{S1}$$

where we can assume that the interdiffusion coefficient $(\widetilde{D})$ does not vary significantly over the small composition range of the phase of interest.

Using standard thermodynamic relation $dC_B = \left(\frac{\overline{V}_A}{V_m^2}\right) dN_B$ [1] in Fick's first law [4], $\widetilde{D}_{int}^{\beta}$ can be related to the interdiffusion flux of component B as

$$\tilde{J}_B = -\widetilde{D}\frac{\partial C_B}{\partial x} = -\widetilde{D}\frac{\overline{V}_A}{V_m^2}\frac{\partial N_B}{\partial x} \tag{S2a}$$

$$\tilde{J}_B = -\widetilde{D}\frac{\overline{V}_A^{\beta}}{\left(V_m^{\beta}\right)^2}\frac{\Delta N_B^{\beta}}{\Delta x^{\beta}} \tag{S2b}$$

where $\Delta N_B^{\beta} = N_B^{\beta_2} - N_B^{\beta_1}$ is the narrow homogeneity range of the β phase and $\Delta x^{\beta} = x^{\beta_2} - x^{\beta_1}$ is the thickness of the β phase. Note here that for the phase with narrow homogeneity range, the unknown variation of the slope with the composition (or hence the location parameter *x*) is considered as linear, *i.e.*, $(\Delta N/\Delta x)$. Further, using $\overline{V}_B^{\beta}\tilde{J}_B^{\beta} = -\overline{V}_A^{\beta}\tilde{J}_A^{\beta}$ from Equation 6, and utilizing the definition of the integrated interdiffusion coefficient (Equation S1), we can relate it with the interdiffusion fluxes of component B and A as

$$\widetilde{D}_{int}^{\beta} = -\frac{\left(V_m^{\beta}\right)^2}{\overline{V}_A^{\beta}} \Delta x^{\beta} \tilde{J}_B^{\beta} = \frac{\left(V_m^{\beta}\right)^2}{\overline{V}_B^{\beta}} \Delta x^{\beta} \tilde{J}_A^{\beta} \tag{S3}$$

Note that in the case of normal downhill diffusion, the composition profile is plotted such that the component B diffuse from right to left and component A diffuse from left to right. In such a situation, the interdiffusion flux at one particular composition located at a particular location of the diffusion couple after annealing for time $t$, $\tilde{J}_B$ will have a negative sign and $\tilde{J}_A$ will have a positive sign leading equal and positive value of $\tilde{D}_{int}^{\beta}$ irrespective of the composition profile considered for the estimation. With respect the concentration profiles of components B and A, following the Equations 20, we have

$$\tilde{J}_B^{\beta} = \tilde{J}_B(C_B^{\beta}) = -\left(\frac{C_B^+ - C_B^-}{2t}\right)\left[\left(1 - Y_{C_B}^{\beta}\right)\int_{x^{-\infty}}^{x^*} Y_{C_B}\, dx + Y_{C_B}^{\beta}\int_{x^*}^{x^{+\infty}}(1 - Y_{C_B})\, dx\right] \quad \text{(S4a)}$$

$$\tilde{J}_A^{\beta} = \tilde{J}_A(C_A^{\beta}) = -\left(\frac{C_A^- - C_A^+}{2t}\right)\left[\left(1 - Y_{C_A}^{\beta}\right)\int_{x^{+\infty}}^{x^*} Y_{C_A}\, dx + Y_{C_A}^{\beta}\int_{x^*}^{x^{-\infty}}(1 - Y_{C_A})\, dx\right] \quad \text{(S4b)}$$

where $Y_{C_B}^{\beta} = \frac{C_B^{\beta} - C_B^-}{C_B^+ - C_B^-}$ and $Y_{C_A}^{\beta} = \frac{C_A^{\beta} - C_A^+}{C_A^- - C_A^+}$.

The term inside square bracket is separated into 3 parts in the interdiffusion zone as the thickness related to the phase of interest and the other two parts for the interdiffusion zone before and after that:

$$\tilde{J}_B^{\beta} = -\left(\frac{C_B^+ - C_B^-}{2t}\right)\left[\left(1 - Y_{C_B}^{\beta}\right)\int_{x^{-\infty}}^{x^{\beta_1}} Y_{C_B}\, dx + Y_{C_B}^{\beta}\int_{x^{\beta_1}}^{x^{\beta_2}}(1 - Y_{C_B})\, dx + Y_{C_B}^{\beta}\int_{x^{\beta_2}}^{x^{+\infty}}(1 - Y_{C_B})\, dx\right] \quad \text{(S5a)}$$

$$\tilde{J}_A^{\beta} = -\left(\frac{C_A^- - C_A^+}{2t}\right)\left[\left(1 - Y_{C_A}^{\beta}\right)\int_{x^{+\infty}}^{x^{\beta_2}} Y_{C_A}\, dx + Y_{C_A}^{\beta}\int_{x^{\beta_2}}^{x^{\beta_1}}(1 - Y_{C_A})\, dx + Y_{C_A}^{\beta}\int_{x^{\beta_1}}^{x^{-\infty}}(1 - Y_{C_A})\, dx\right] \quad \text{(S5b)}$$

In the phase of interest $Y_{C_B}^{\beta}$ (or $Y_{C_A}^{\beta}$) is constant because of the growth of the phase with very narrow homogeneity range, *i.e.*, with almost a fixed composition $C_B^{\beta}$ (or $C_A^{\beta}$). Therefore, after rearranging, we can write

$$\tilde{J}_B^{\beta} = -\left(\frac{C_B^+ - C_B^-}{2t}\right)\left[Y_{C_B}^{\beta}\left(1 - Y_{C_B}^{\beta}\right)\Delta x^{\beta} + \left(1 - Y_{C_B}^{\beta}\right)\int_{x^{-\infty}}^{x^{\beta_1}} Y_{C_B}\, dx + Y_{C_B}^{\beta}\int_{x^{\beta_2}}^{x^{+\infty}}(1 - Y_{C_B})\, dx\right] \quad \text{(S6a)}$$

$$\tilde{J}_A^{\beta} = \left(\frac{C_A^- - C_A^+}{2t}\right)\left[Y_{C_A}^{\beta}(1 - Y_{C_A}^{\beta})\Delta x^{\beta} + \left(1 - Y_{C_A}^{\beta}\right)\int_{x^{\beta_2}}^{x^{+\infty}} Y_{C_A}\, dx + Y_{C_A}^{\beta}\int_{x^{-\infty}}^{x^{\beta_1}}(1 - Y_{C_A})\, dx\right] \quad \text{(S6b)}$$

where $\Delta x^{\beta} = x^{\beta_2} - x^{\beta_1}$ is the thickness of the β phase. It should be noted there that the minus sign in $\tilde{J}_A^{\beta}$ is omitted because of changing the limits of integration. Therefore, from Equation S3, the integrated diffusion coefficient with respect to $C_B$ and $C_A$ can be expressed as

$$\tilde{D}_{int}^{\beta} = -\frac{\left(V_m^{\beta}\right)^2}{\overline{V}_A^{\beta}}\Delta x^{\beta}\tilde{J}_B^{\beta} = \frac{\left(V_m^{\beta}\right)^2}{\overline{V}_A^{\beta}}\Delta x^{\beta}\left(\frac{C_B^+ - C_B^-}{2t}\right)\left[Y_{C_B}^{\beta}\left(1 - Y_{C_B}^{\beta}\right)\Delta x^{\beta} + \left(1 - Y_{C_B}^{\beta}\right)\int_{x^{-\infty}}^{x^{\beta_1}} Y_{C_B}\, dx + Y_{C_B}^{\beta}\int_{x^{\beta_2}}^{x^{+\infty}}(1 - Y_{C_B})\, dx\right] \quad \text{(S7a)}$$

$$\widetilde{D}^{\beta}_{int} = \frac{\left(V_m^{\beta}\right)^2}{\bar{V}_B^{\beta}} \Delta x^{\beta} \tilde{J}_A^{\beta} = \frac{\left(V_m^{\beta}\right)^2}{\bar{V}_B^{\beta}} \Delta x^{\beta} \left(\frac{C_A^- - C_A^+}{2t}\right) \left[Y_{C_A}^{\beta}(1 - Y_{C_A}^{\beta})\Delta x^{\beta} + \left(1 - Y_{C_A}^{\beta}\right) \int_{x^{\beta_2}}^{x^{+\infty}} Y_{C_A} \, dx + Y_{C_A}^{\beta} \int_{x^{-\infty}}^{x^{\beta_1}} (1 - Y_{C_A}) \, dx\right] \tag{S7b}$$

Further, expanding $Y_{C_B}$ and $Y_{C_A}$ (from Equation S4) with respect to $C_B$ and $C_A$, we get

$$\widetilde{D}^{\beta}_{int} = \frac{\left(V_m^{\beta}\right)^2}{\bar{V}_A^{\beta}} \frac{\left(C_B^{\beta} - C_B^-\right)\left(C_B^+ - C_B^{\beta}\right)}{\left(C_B^+ - C_B^-\right)} \frac{(\Delta x^{\beta})^2}{2t}$$

$$+ \frac{\left(V_m^{\beta}\right)^2}{\bar{V}_A^{\beta}} \frac{\Delta x^{\beta}}{2t} \left[\frac{C_B^+ - C_B^{\beta}}{C_B^+ - C_B^-} \int_{x^{-\infty}}^{x^{\beta_1}} (C_B - C_B^-) \, dx + \frac{C_B^{\beta} - C_B^-}{C_B^+ - C_B^-} \int_{x^{\beta_2}}^{x^{+\infty}} (C_B^+ - C_B) \, dx\right]$$

(S8a)

$$\widetilde{D}^{\beta}_{int} = \frac{\left(V_m^{\beta}\right)^2}{\bar{V}_B^{\beta}} \frac{\left(C_A^{\beta} - C_A^+\right)\left(C_A^- - C_A^{\beta}\right)}{\left(C_A^- - C_A^+\right)} \frac{(\Delta x^{\beta})^2}{2t}$$

$$+ \frac{\left(V_m^{\beta}\right)^2}{\bar{V}_B^{\beta}} \frac{\Delta x^{\beta}}{2t} \left[\frac{C_A^- - C_A^{\beta}}{C_A^- - C_A^+} \int_{x^{\beta_2}}^{x^{+\infty}} (C_A - C_A^+) \, dx + \frac{C_A^{\beta} - C_A^+}{C_A^- - C_A^+} \int_{x^{-\infty}}^{x^{\beta_1}} (C_A^- - C_A) \, dx\right]$$

(S8b)

Note that $C_A + C_B = \frac{1}{V_m}$ and $C_B^+ > C_B^-$, $C_A^- > C_A^+$.

Therefore, Equations S7 or S8 are relations for the estimation of $\widetilde{D}^{\beta}_{int}$ with respect to $Y_{C_B}$ (and $Y_{C_A}$) or $C_B$ (and $C_A$), which was not available earlier. Previously, Wagner [10] derived the relation with respect to $Y_{N_B}$ or $N_B$ considering non–ideal variation of the molar volume, which can be expressed as [1]

The interdiffusion fluxes from the composition profiles of components B and A are

$$\tilde{J}_B^{\beta} = -\frac{\bar{V}_A^{\beta}}{V_m^{\beta}} \left(\frac{N_B^+ - N_B^-}{2t}\right) \left[\frac{Y_{N_B}^{\beta}(1 - Y_{N_B}^{\beta})}{V_m^{\beta}} \Delta x^{\beta} + \left(1 - Y_{N_B}^{\beta}\right) \int_{x^{-\infty}}^{x^{\beta_1}} \frac{Y_{N_B}}{V_m} \, dx + Y_{N_B}^{\beta} \int_{x^{\beta_2}}^{x^{+\infty}} \frac{(1 - Y_{N_B})}{V_m} \, dx\right] \tag{S9a}$$

$$\tilde{J}_A^{\beta} = \frac{\bar{V}_B^{\beta}}{V_m^{\beta}} \left(\frac{N_A^- - N_A^+}{2t}\right) \left[\frac{Y_{N_A}^{\beta}(1 - Y_{N_A}^{\beta})}{V_m^{\beta}} \Delta x^{\beta} + \left(1 - Y_{N_A}^{\beta}\right) \int_{x^{\beta_2}}^{x^{+\infty}} \frac{Y_{N_A}}{V_m} \, dx + Y_{N_A}^{\beta} \int_{x^{-\infty}}^{x^{\beta_1}} \frac{(1 - Y_{N_A})}{V_m} \, dx\right] \tag{S9b}$$

where $Y_{N_B}^{\beta} = \frac{N_B^{\beta} - N_B^-}{N_B^+ - N_B^-}$ and $Y_{N_A}^{\beta} = \frac{N_A^{\beta} - N_A^+}{N_A^- - N_A^+}$.

Therefore, from Equation S3, the $\widetilde{D}^{\beta}_{int}$ with respect to $N_B$ and $N_A$ can be expressed as

$$\widetilde{D}_{int}^{\beta} = -\frac{\left(V_m^{\beta}\right)^2}{\overline{V}_A^{\beta}}\Delta x^{\beta} \tilde{J}_B^{\beta} = V_m^{\beta}\Delta x^{\beta}\left(\frac{N_B^+ - N_B^-}{2t}\right)\left[\frac{Y_{N_B}^{\beta}\left(1-Y_{N_B}^{\beta}\right)}{V_m^{\beta}}\Delta x^{\beta} + \left(1 - Y_{N_B}^{\beta}\right)\int_{x^{-\infty}}^{x^{\beta_1}}\frac{Y_{N_B}}{V_m}dx +$$

$$Y_{N_B}^{\beta}\int_{x^{\beta_2}}^{x^{+\infty}}\frac{(1-Y_{N_B})}{V_m}dx\right] \tag{S10a}$$

$$\widetilde{D}_{int}^{\beta} = \frac{\left(V_m^{\beta}\right)^2}{\overline{V}_B^{\beta}}\Delta x^{\beta} \tilde{J}_A^{\beta} = V_m^{\beta}\Delta x^{\beta}\left(\frac{N_A^- - N_A^+}{2t}\right)\left[\frac{Y_{N_A}^{\beta}\left(1-Y_{N_A}^{\beta}\right)}{V_m^{\beta}}\Delta x^{\beta} + \left(1 - Y_{N_A}^{\beta}\right)\int_{x^{\beta_2}}^{x^{+\infty}}\frac{Y_{N_A}}{V_m}dx +$$

$$Y_{N_A}^{\beta}\int_{x^{-\infty}}^{x^{\beta_1}}\frac{(1-Y_{N_A})}{V_m}dx\right] \tag{S10b}$$

Further, expanding $Y_{N_B}$ and $Y_{N_A}$ (from Equation S9) with respect to $N_B$ and $N_A$, we get

$$\widetilde{D}_{int}^{\beta} = \frac{\left(N_B^{\beta} - N_B^-\right)\left(N_B^+ - N_B^{\beta}\right)}{\left(N_B^+ - N_B^-\right)}\frac{(\Delta x^{\beta})^2}{2t}$$

$$+ \frac{V_m^{\beta}\Delta x^{\beta}}{2t}\left[\frac{\left(N_B^+ - N_B^{\beta}\right)}{\left(N_B^+ - N_B^-\right)}\int_{x^{-\infty}}^{x^{\beta_1}}\frac{(N_B - N_B^-)}{V_m}dx + \frac{\left(N_B^{\beta} - N_B^-\right)}{\left(N_B^+ - N_B^-\right)}\int_{x^{\beta_2}}^{x^{+\infty}}\frac{(N_B^+ - N_B)}{V_m}dx\right]$$

(S11a)

$$\widetilde{D}_{int}^{\beta} = \frac{\left(N_A^{\beta} - N_A^+\right)\left(N_A^- - N_A^{\beta}\right)}{\left(N_A^- - N_A^+\right)}\frac{(\Delta x^{\beta})^2}{2t}$$

$$+ \frac{V_m^{\beta}\Delta x^{\beta}}{2t}\left[\frac{\left(N_A^- - N_A^{\beta}\right)}{\left(N_A^- - N_A^+\right)}\int_{x^{\beta_2}}^{x^{+\infty}}\frac{(N_A - N_A^+)}{V_m}dx + \frac{\left(N_A^{\beta} - N_A^+\right)}{\left(N_A^- - N_A^+\right)}\int_{x^{-\infty}}^{x^{\beta_1}}\frac{(N_A^- - N_A)}{V_m}dx\right]$$

(S11b)

Note that $N_A + N_B = 1$ and $N_B^+ > N_B^-$, $N_A^- > N_A^+$. Equations S10 or S11 was derived by Wagner [10], which is expressed with respect to $Y_{N_B}$ (and $Y_{N_A}$) or $N_B$ (and $N_A$).

In an intermetallic compound with narrow homogeneity range, we cannot estimate the composition or concentration gradients. Even we do not know the partial molar volumes of the components in a phase. Therefore, instead of following the Equation 29b, we can estimate the ratio of the tracer diffusion coefficients by neglecting the vacancy wind effect following Equations 29a, c and 32 as

$$\frac{D_B^*}{D_A^*} = \left[\frac{C_B^+\int_{x^{-\infty}}^{x^K}Y_{C_B}dx - C_B^-\int_{x^K}^{x^{+\infty}}(1-Y_{C_B})dx}{-C_A^-\int_{x^{+\infty}}^{x^K}Y_{C_A}dx + C_A^+\int_{x^K}^{x^{-\infty}}(1-Y_{C_A})dx}\right] \tag{S12}$$

Note here that the contribution of the vacancy wind effect does not contribute very significantly in most of the systems and the difference in estimated data could fall within the limit of experimental error [1]. The same relation with respect to $Y_{N_B}$ and $N_B$ (following Equations 30 and 32) can be expressed as

$$\frac{D_B^*}{D_A^*} = \left[\frac{N_B^+ \int_{x-\infty}^{x^K} \frac{Y_{N_B}}{V_m} dx - N_B^- \int_{x^K}^{x+\infty} \frac{(1-Y_{N_B})}{V_m} dx}{-N_A^+ \int_{x-\infty}^{x^K} \frac{Y_{N_B}}{V_m} dx + N_A^- \int_{x^K}^{x+\infty} \frac{(1-Y_{N_B})}{V_m} dx}\right] \quad (S13)$$

Since these relations are free from partial molar volume terms, there data can be estimated straightforwardly.

**Estimation of the integrated diffusion coefficients following different methods in the Cu–Sn system**

The interdiffusion zone is shown in Figure 4a. The average thicknesses of the phases are $\Delta x^{Cu_3Sn} = 3.5$ μm and $\Delta x^{Cu_6Sn_5} = 13$ μm. The couple was annealed for 81 hrs and therefore $2t = 2\times81\times3600$ s. Marker plane in Cu$_6$Sn$_5$ phase is found at a distance of 7 μm from Cu$_3$Sn/Cu$_6$Sn$_5$ interface. The molar volumes of the phases are $V_m^{Cu_3Sn} = 8.59\times10^{-6}$ and $V_m^{Cu_6Sn_5} = 10.59\times10^{-6}$ m³/mol.

The composition profile with respect to component Sn and Cu is shown in the Figure S1.

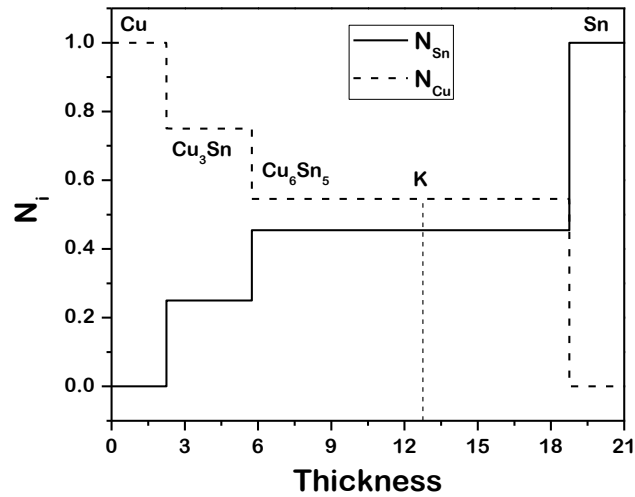

Figure S1: The average composition profiles of Cu and Sn in the Cu–Sn diffusion couple annealed at 200 °C for 81 hrs. The micrograph is shown in Figure 4a.

**S2. Estimation with respect to the composition normalized variable following the Wagner method:**

Composition normalized variables with respect to component Sn and Cu are

$$Y_{N_{Sn}}^{Cu_3Sn} = \frac{N_{Sn}^{Cu_3Sn} - N_{Sn}^-}{N_{Sn}^+ - N_{Sn}^-} = \frac{\frac{1}{4} - 0}{1-0} = \frac{1}{4}, \quad Y_{N_{Sn}}^{Cu_6Sn_5} = \frac{N_{Sn}^{Cu_6Sn_5} - N_{Sn}^-}{N_{Sn}^+ - N_{Sn}^-} = \frac{\frac{5}{11} - 0}{1-0} = \frac{5}{11}$$

$$Y_{N_{Cu}}^{Cu_3Sn} = \frac{N_{Cu}^{Cu_3Sn} - N_{Cu}^+}{N_{Cu}^- - N_{Cu}^+} = \frac{\frac{3}{4} - 0}{1-0} = \frac{3}{4}, \quad Y_{N_{Cu}}^{Cu_6Sn_5} = \frac{N_{Cu}^{Cu_6Sn_5} - N_{Cu}^+}{N_{Cu}^- - N_{Cu}^+} = \frac{\frac{6}{11} - 0}{1-0} = \frac{6}{11}$$

**S2.1 Estimation in the Cu₃Sn phase**

For the Cu₃Sn phase, since there is no phase in the interdiffusion zone between Cu and the phase of interest, the second term inside the bracket in Equation S9a becomes zero and we can write the interdiffusion flux from the composition profile of component Sn as

$$\tilde{J}_{Sn}^{Cu_3Sn} = -\frac{\bar{V}_{Cu}^{Cu_3Sn}}{V_m^{Cu_3Sn}}\left(\frac{N_{Sn}^+ - N_{Sn}^-}{2t}\right)\left[\frac{Y_{N_{Sn}}^{Cu_3Sn}\left(1-Y_{N_{Sn}}^{Cu_3Sn}\right)}{V_m^{Cu_3Sn}}\Delta x^{Cu_3Sn} + 0 + Y_{N_{Sn}}^{Cu_3Sn}\frac{\left(1-Y_{N_{Sn}}^{Cu_6Sn_5}\right)}{V_m^{Cu_6Sn_5}}\Delta x^{Cu_6Sn_5}\right]$$

$$\tilde{J}_{Sn}^{Cu_3Sn} = -\frac{\bar{V}_{Cu}^{Cu_3Sn}}{V_m^{Cu_3Sn}}\left(\frac{1-0}{2\times 81\times 3600}\right)\left[\frac{\frac{1}{4}\left(1-\frac{1}{4}\right)}{8.59\times 10^{-6}}3.5\times 10^{-6} + 0 + \frac{1}{4}\frac{\left(1-\frac{5}{11}\right)}{10.59\times 10^{-6}}13\times 10^{-6}\right]$$

$$\tilde{J}_{Sn}^{Cu_3Sn} = -\frac{\bar{V}_{Cu}^{Cu_3Sn}}{V_m^{Cu_3Sn}}\times 4.18\times 10^{-7} \text{ mol/m}^2.s$$

Following Equation S3,

$$\tilde{D}_{int}^{Cu_3Sn} = \frac{\left(V_m^{Cu_3Sn}\right)^2}{\bar{V}_{Cu}^{Cu_3Sn}}\Delta x^{Cu_3Sn}\left(-\tilde{J}_{Sn}^{Cu_3Sn}\right)$$

$$\tilde{D}_{int}^{Cu_3Sn} = \frac{\left(V_m^{Cu_3Sn}\right)^2}{\bar{V}_{Sn}^{Cu_3Sn}}\Delta x^{Cu_3Sn}\left(\frac{\bar{V}_{Sn}^{Cu_3Sn}}{V_m^{Cu_3Sn}}\times 4.18\times 10^{-7}\right)$$

$$\tilde{D}_{int}^{Cu_3Sn} = V_m^{Cu_3Sn}\Delta x^{Cu_3Sn}(4.18\times 10^{-7}) = 8.59\times 10^{-6}\times 3.5\times 10^{-6}\times(4.18\times 10^{-7})$$

$$\tilde{D}_{int}^{Cu_3Sn} = 1.26\times 10^{-17} \text{ m}^2/s$$

Following Equation S9b, since there is Cu₆Sn₅ phase in the interdiffusion zone between Sn and Cu₃Sn, and no phase is between Cu₃Sn and Cu, we can write the interdiffusion flux with respect to component Cu as

$$\tilde{J}_{Cu}^{Cu_3Sn} = \frac{\bar{V}_{Sn}^{Cu_3Sn}}{V_m^{Cu_3Sn}}\left(\frac{N_{Cu}^- - N_{Cu}^+}{2t}\right)\left[\frac{Y_{N_{Cu}}^{Cu_3Sn}\left(1-Y_{N_{Cu}}^{Cu_3Sn}\right)}{V_m^{Cu_3Sn}}\Delta x^{Cu_3Sn} + \left(1-Y_{N_{Cu}}^{Cu_3Sn}\right)\frac{Y_{N_{Cu}}^{Cu_6Sn_5}}{V_m^{Cu_6Sn_5}}\Delta x^{Cu_6Sn_5} + 0\right]$$

$$\tilde{J}_{Cu}^{Cu_3Sn} = \frac{\bar{V}_{Sn}^{Cu_3Sn}}{V_m^{Cu_3Sn}}\left(\frac{1-0}{2\times 81\times 3600}\right)\left[\frac{\frac{3}{4}\left(1-\frac{3}{4}\right)}{8.59\times 10^{-6}}3.5\times 10^{-6} + \left(1-\frac{3}{4}\right)\frac{\frac{6}{11}}{10.59\times 10^{-6}}13\times 10^{-6} + 0\right]$$

$$\tilde{J}_{Cu}^{Cu_3Sn} = \frac{\bar{V}_{Sn}^{Cu_3Sn}}{V_m^{Cu_3Sn}} \times 4.18 \times 10^{-7} \ mol/m^2.s$$

Following Equation S3,

$$\tilde{D}_{int}^{Cu_3Sn} = \frac{(V_m^{Cu_3Sn})^2}{\bar{V}_{Cu}^{Cu_3Sn}} \Delta x^{Cu_3Sn} (\tilde{J}_{Cu}^{Cu_3Sn})$$

$$\tilde{D}_{int}^{Cu_3Sn} = \frac{(V_m^{Cu_3Sn})^2}{\bar{V}_{Sn}^{Cu_3Sn}} \Delta x^{Cu_3Sn} \left( \frac{\bar{V}_{Sn}^{Cu_3Sn}}{V_m^{Cu_3Sn}} \times 4.18 \times 10^{-7} \right)$$

$$\tilde{D}_{int}^{Cu_3Sn} = V_m^{Cu_3Sn} \Delta x^{Cu_3Sn} (4.18 \times 10^{-7}) = 8.59 \times 10^{-6} \times 3.5 \times 10^{-6} \times (4.18 \times 10^{-7})$$

$$\tilde{D}_{int}^{Cu_3Sn} = 1.26 \times 10^{-17} \ m^2/s$$

Therefore, as it should be, the exactly same value is estimated following the composition profiles of Sn and Cu. Most importantly, one has to make sure that Equation 6 is fulfilled. This is indeed fulfilled since

$$\bar{V}_{Cu}^{Cu_3Sn} \tilde{J}_{Cu}^{Cu_3Sn} + \bar{V}_{Sn}^{Cu_3Sn} \tilde{J}_{Sn}^{Cu_3Sn} = \bar{V}_{Cu}^{Cu_3Sn} \frac{\bar{V}_{Sn}^{Cu_3Sn}}{V_m^{Cu_3Sn}} \times 4.18 \times 10^{-7} - \bar{V}_{Sn}^{Cu_3Sn} \frac{\bar{V}_{Cu}^{Cu_3Sn}}{V_m^{Cu_3Sn}} \times 4.18 \times 10^{-7} = 0$$

## S2.2 Estimation in the Cu$_6$Sn$_5$ phase

Following Equation S9a, since there is no phase in the interdiffusion zone between Cu$_6$Sn$_5$ and Sn, we can write the interdiffusion flux with respect to component Sn

$$\tilde{J}_{Sn}^{Cu_6Sn_5} = -\frac{\bar{V}_{Cu}^{Cu_6Sn_5}}{V_m^{Cu_6Sn_5}} \left( \frac{N_{Sn}^+ - N_{Sn}^-}{2t} \right) \left[ \frac{Y_{N_{Sn}}^{Cu_6Sn_5}(1-Y_{N_{Sn}}^{Cu_6Sn_5})}{V_m^{Cu_6Sn_5}} \Delta x^{Cu_6Sn_5} + (1-Y_{N_{Sn}}^{Cu_6Sn_5}) \frac{Y_{N_{Sn}}^{Cu_3Sn}}{V_m^{Cu_3Sn}} \Delta x^{Cu_3Sn} + 0 \right]$$

$$\tilde{J}_{Sn}^{Cu_6Sn_5} = -\frac{\bar{V}_{Cu}^{Cu_6Sn_5}}{V_m^{Cu_6Sn_5}} \left( \frac{1-0}{2 \times 81 \times 3600} \right) \left[ \frac{\frac{5}{11}(1-\frac{5}{11})}{10.59 \times 10^{-6}} 13 \times 10^{-6} + \left(1-\frac{5}{11}\right) \frac{\frac{1}{4}}{8.59 \times 10^{-6}} 3.5 \times 10^{-6} + 0 \right]$$

$$\tilde{J}_{Sn}^{Cu_6Sn_5} = -\frac{\bar{V}_{Cu}^{Cu_6Sn_5}}{V_m^{Cu_6Sn_5}} 6.17 \times 10^{-7} \ mol/m^2.s$$

Following Equation S3,

$$\tilde{D}_{int}^{Cu_6Sn_5} = \frac{(V_m^{Cu_6Sn_5})^2}{\bar{V}_{Cu}^{Cu_6Sn_5}} \Delta x^{Cu_6Sn_5} (-\tilde{J}_{Sn}^{Cu_6Sn_5})$$

$$\tilde{D}_{int}^{Cu_6Sn_5} = \frac{(V_m^{Cu_6Sn_5})^2}{\bar{V}_{Cu}^{Cu_6Sn_5}} \Delta x^{Cu_6Sn_5} \left( \frac{\bar{V}_{Cu}^{Cu_6Sn_5}}{V_m^{Cu_6Sn_5}} 6.17 \times 10^{-7} \right)$$

$$\tilde{D}_{int}^{Cu_6Sn_5} = V_m^{Cu_6Sn_5} \Delta x^{Cu_6Sn_5} (6.17 \times 10^{-7}) = 10.59 \times 10^{-6} \times 13 \times 10^{-6} \times (6.17 \times 10^{-7})$$

$$\tilde{D}_{int}^{Cu_6Sn_5} = 8.49 \times 10^{-17} \ m^2/s$$

Following Equation S9b, since there is no phase in the interdiffusion zone between Sn and $Cu_6Sn_5$, and $Cu_3Sn$ phase is between $Cu_6Sn_5$ and Cu, we can write the interdiffusion flux with respect to component Cu as

$$\tilde{J}_{Cu}^{Cu_6Sn_5} = \frac{\bar{V}_{Sn}^{Cu_6Sn_5}}{V_m^{Cu_6Sn_5}} \left(\frac{N_{Cu}^- - N_{Cu}^+}{2t}\right) \left[\frac{Y_{N_{Cu}}^{Cu_6Sn_5}\left(1-Y_{N_{Cu}}^{Cu_6Sn_5}\right)}{V_m^{Cu_6Sn_5}} \Delta x^{Cu_6Sn_5} + 0 + Y_{N_{Cu}}^{Cu_6Sn_5} \frac{\left(1-Y_{N_{Cu}}^{Cu_3Sn}\right)}{V_m^{Cu_3Sn}} \Delta x^{Cu_3Sn}\right]$$

$$\tilde{J}_{Cu}^{Cu_6Sn_5} = \frac{\bar{V}_{Sn}^{Cu_6Sn_5}}{V_m^{Cu_6Sn_5}} \left(\frac{1-0}{2\times 81\times 3600}\right) \left[\frac{\frac{6}{11}\left(1-\frac{6}{11}\right)}{10.59\times 10^{-6}} 13\times 10^{-6} + 0 + \frac{6}{11}\frac{\left(1-\frac{3}{4}\right)}{8.59\times 10^{-6}} 3.5\times 10^{-6}\right]$$

$$\tilde{J}_{Cu}^{Cu_6Sn_5} = \frac{\bar{V}_{Sn}^{Cu_6Sn_5}}{V_m^{Cu_6Sn_5}} 6.17\times 10^{-7} \ mol/m^2.s$$

Following Equation S3,

$$\tilde{D}_{int}^{Cu_6Sn_5} = \frac{\left(V_m^{Cu_6Sn_5}\right)^2}{\bar{V}_{Sn}^{Cu_6Sn_5}} \Delta x^{Cu_6Sn_5} \left(\tilde{J}_{Cu}^{Cu_6Sn_5}\right)$$

$$\tilde{D}_{int}^{Cu_6Sn_5} = \frac{\left(V_m^{Cu_6Sn_5}\right)^2}{\bar{V}_{Sn}^{Cu_6Sn_5}} \Delta x^{Cu_6Sn_5} \left(\frac{\bar{V}_{Sn}^{Cu_6Sn_5}}{V_m^{Cu_6Sn_5}} 6.17\times 10^{-7}\right)$$

$$\tilde{D}_{int}^{Cu_6Sn_5} = V_m^{Cu_6Sn_5} \Delta x^{Cu_6Sn_5} (6.17\times 10^{-7}) = 10.59\times 10^{-6}\times 13\times 10^{-6}\times (6.17\times 10^{-7})$$

$$\tilde{D}_{int}^{Cu_6Sn_5} = 8.49\times 10^{-17} \ m^2/s$$

Therefore, again we have the same values when estimated with respect to component Sn and Cu. We can also verify the Equation 6 following

$$\bar{V}_{Cu}^{Cu_6Sn_5} \tilde{J}_{Cu}^{Cu_6Sn_5} + \bar{V}_{Sn}^{Cu_6Sn_5} \tilde{J}_{Sn}^{Cu_6Sn_5} = \bar{V}_{Cu}^{Cu_6Sn_5} \frac{\bar{V}_{Sn}^{Cu_6Sn_5}}{V_m^{Cu_6Sn_5}} 6.17\times 10^{-7} - \bar{V}_{Sn}^{Cu_6Sn_5} \frac{\bar{V}_{Cu}^{Cu_6Sn_5}}{V_m^{Cu_6Sn_5}} 6.17\times 10^{-7} = 0$$

It is to be noted there that although the partial molar volume terms are unknown, we could still verify the condition in Equation 6 is indeed fulfill. However, the same is not true with respect to the concentration normalized variable, as shown in the next section.

**S3.   Estimation with respect to the concentration normalized variable following the relations derived in the present work:**

Since the partial molar volumes of components in the $\beta$ phase are unknown, it is evident from Equations S7 or S8 that we cannot estimate $\tilde{D}_{int}^\beta$ directly with respect to $Y_{C_B}$ (and $Y_{C_A}$) or $C_B$ (and $C_A$). To facilitate the discussion on one of the important points as discussed in the manuscript, we estimate data for both the actual and the ideal molar volume of the phase of interest β.

**S3.1   Estimation of the data considering the actual molar volume variation**

For the actual $V_m$ of phases, we can write

$$Y_{C_{Sn}}^{Cu_3Sn} = \frac{C_{Sn}^{Cu_3Sn} - C_{Sn}^-}{C_{Sn}^+ - C_{Sn}^-} = \frac{\left(\frac{1}{4}\right)/(8.59\times10^{-6}) - 0}{1/(16.24\times10^{-6}) - 0} = \frac{1\times16.24}{4\times8.59}$$

$$Y_{C_{Sn}}^{Cu_6Sn_5} = \frac{C_{Sn}^{Cu_6Sn_5} - C_{Sn}^-}{C_{Sn}^+ - C_{Sn}^-} = \frac{\left(\frac{5}{11}\right)/(10.59\times10^{-6}) - 0}{1/(16.24\times10^{-6}) - 0} = \frac{5\times16.24}{11\times10.59}$$

$$Y_{C_{Cu}}^{Cu_3Sn} = \frac{C_{Cu}^{Cu_3Sn} - C_{Cu}^+}{C_{Cu}^- - C_{Cu}^+} = \frac{\left(\frac{3}{4}\right)/(8.59\times10^{-6}) - 0}{1/(7.12\times10^{-6}) - 0} = \frac{3\times7.12}{4\times8.59}$$

$$Y_{C_{Cu}}^{Cu_6Sn_5} = \frac{C_{Cu}^{Cu_6Sn_5} - C_{Cu}^+}{C_{Cu}^- - C_{Cu}^+} = \frac{\left(\frac{6}{11}\right)/(10.59\times10^{-6}) - 0}{1/(7.12\times10^{-6}) - 0} = \frac{6\times7.12}{11\times10.59}$$

**S3.1.1 Estimation in the Cu₃Sn phase**

Following Equation S6a, we can write the interdiffusion flux with respect to component Sn as

$$\tilde{J}_{Sn}^{Cu_3Sn} = -\left(\frac{C_{Sn}^+ - C_{Sn}^-}{2t}\right)\left[Y_{C_{Sn}}^{Cu_3Sn}\left(1 - Y_{C_{Sn}}^{Cu_3Sn}\right)\Delta x^{Cu_3Sn} + 0 + Y_{C_{Sn}}^{Cu_3Sn}\left(1 - Y_{C_{Sn}}^{Cu_6Sn_5}\right)\Delta x^{Cu_6Sn_5}\right]$$

$$\tilde{J}_{Sn}^{Cu_3Sn} = -\left(\frac{\frac{1}{16.24\times10^{-6}}}{2\times81\times3600}\right)\left[\frac{16.24}{4\times8.59}\left(1 - \frac{16.24}{4\times8.59}\right)3.5\times10^{-6} + \frac{16.24}{4\times8.59}\left(1 - \frac{5\times16.24}{11\times10.59}\right)13\times10^{-6}\right]$$

$$\tilde{J}_{Sn}^{Cu_3Sn} = -2.89\times10^{-7}\ mol/m^2.s$$

Following Equation S3,

$$\widetilde{D}_{int}^{Cu_3Sn} = \frac{\left(V_m^{Cu_3Sn}\right)^2}{\overline{V}_{Cu}^{Cu_3Sn}}\Delta x^{Cu_3Sn}\left(-\tilde{J}_{Sn}^{Cu_3Sn}\right)$$

$$\widetilde{D}_{int}^{Cu_3Sn} = \frac{\left(V_m^{Cu_3Sn}\right)^2}{\overline{V}_{Cu}^{Cu_3Sn}}\Delta x^{Cu_3Sn}(2.89\times10^{-7})$$

$$\widetilde{D}_{int}^{Cu_3Sn} = \frac{V_m^{Cu_3Sn}}{\overline{V}_{Cu}^{Cu_3Sn}}8.59\times10^{-6}\times3.5\times10^{-6}\times(2.89\times10^{-7})$$

$$\widetilde{D}_{int}^{Cu_3Sn} = \frac{V_m^{Cu_3Sn}}{\overline{V}_{Cu}^{Cu_3Sn}}(8.69\times10^{-18})\ m^2/s.$$

Since partial molar volumes are not known it is a common practice to consider the partial molar volumes as equal to the molar volume in a phase with narrow homogeneity range. Therefore, the estimated value would be

$$\widetilde{D}_{int}^{Cu_3Sn} = 8.69\times10^{-18}\ m^2/s$$

However, a major problem is faced with this assumption (*i.e.*, considering $\bar{V}_i^\beta = V_m^\beta$), when the same data are estimated with respect to the composition profile of another component. Following Equation S6b, we can write the interdiffusion flux with respect to component Cu as

$$\tilde{J}_{Cu}^{Cu_3Sn} = \left(\frac{C_{Cu}^- - C_{Cu}^+}{2t}\right)\left[Y_{C_{Cu}}^{Cu_3Sn}(1 - Y_{C_{Cu}}^{Cu_3Sn})\Delta x^{Cu_3Sn} + (1 - Y_{C_{Cu}}^{Cu_3Sn})Y_{C_{Cu}}^{Cu_6Sn_5}\Delta x^{Cu_6Sn_5} + 0\right]$$

$$\tilde{J}_{Cu}^{Cu_3Sn} = \left(\frac{\frac{1}{7.12\times10^{-6}}}{2\times81\times3600}\right)\left[\left(\frac{3\times7.12}{4\times8.59}\right)\left(1 - \frac{3\times7.12}{4\times8.59}\right)3.5\times10^{-6} + \left(1 - \frac{3\times7.12}{4\times8.59}\right)\left(\frac{6\times7.12}{11\times10.59}\right)13\times10^{-6}\right]$$

$$\tilde{J}_{Cu}^{Cu_3Sn} = 6.33\times10^{-7} \text{ mol/m}^2.\text{s}$$

Following Equation S3,

$$\tilde{D}_{int}^{Cu_3Sn} = \frac{(V_m^{Cu_3Sn})^2}{\bar{V}_{Sn}^{Cu_3Sn}}\Delta x^{Cu_3Sn}(\tilde{J}_{Cu}^{Cu_3Sn})$$

$$\tilde{D}_{int}^{Cu_3Sn} = \frac{(V_m^{Cu_3Sn})^2}{\bar{V}_{Sn}^{Cu_3Sn}}\Delta x^{Cu_3Sn}(6.33\times10^{-7})$$

$$\tilde{D}_{int}^{Cu_3Sn} = \frac{V_m^{Cu_3Sn}}{\bar{V}_{Sn}^{Cu_3Sn}}8.59\times10^{-6}\times3.5\times10^{-6}\times(6.33\times10^{-7})$$

$$\tilde{D}_{int}^{Cu_3Sn} = \frac{V_m^{Cu_3Sn}}{\bar{V}_{Sn}^{Cu_3Sn}}(1.90\times10^{-17}) \text{ m}^2/\text{s}$$

Therefore, if we consider again that partial molar volumes are equal to the molar volume of the phase then we have $\tilde{D}_{int}^{Cu_3Sn} = 1.90\times10^{-17}$ m²/s. This lead to different values when the diffusion coefficients are estimated with respect to the component Sn and Cu, which is not acceptable. This is resulted from the fact that the estimated interdiffusion fluxes do not fulfill the condition $\bar{V}_{Cu}^{Cu_3Sn}\tilde{J}_{Cu}^{Cu_3Sn} + \bar{V}_{Sn}^{Cu_3Sn}\tilde{J}_{Sn}^{Cu_3Sn} = 0$, which can be understood (considering the partial molar volumes as the same) from the estimated values of the interdiffusion fluxes.

### S3.1.2 Estimation in the Cu$_6$Sn$_5$ phase

Following Equation S6a, we can write the interdiffusion flux with respect to component Sn as

$$\tilde{J}_{Sn}^{Cu_6Sn_5} = -\left(\frac{C_{Sn}^+ - C_{Sn}^-}{2t}\right)\left[Y_{C_{Sn}}^{Cu_6Sn_5}(1 - Y_{C_{Sn}}^{Cu_6Sn_5})\Delta x^{Cu_6Sn_5} + (1 - Y_{C_{Sn}}^{Cu_6Sn_5})Y_{C_{Sn}}^{Cu_3Sn}\Delta x^{Cu_3Sn} + 0\right]$$

$$\tilde{J}_{Sn}^{Cu_6Sn_5} = -\left(\frac{\frac{1}{16.24\times10^{-6}}}{2\times81\times3600}\right)\left[\frac{5\times16.24}{11\times10.59}\left(1 - \frac{5\times16.24}{11\times10.59}\right)13\times10^{-6} + \left(1 - \frac{5\times16.24}{11\times10.59}\right)\frac{1\times16.24}{4\times8.59}3.5\times10^{-6}\right]$$

$$\tilde{J}_{Sn}^{Cu_6Sn_5} = -3.43\times10^{-7} \text{ mol/m}^2.\text{s}$$

Following Equation S3,

$$\widetilde{D}_{int}^{Cu_6Sn_5} = \frac{\left(V_m^{Cu_6Sn_5}\right)^2}{\overline{V}_{Cu}^{Cu_6Sn_5}} \Delta x^{Cu_6Sn_5}\left(-\widetilde{J}_{Sn}^{Cu_6Sn_5}\right)$$

$$\widetilde{D}_{int}^{Cu_6Sn_5} = \frac{\left(V_m^{Cu_6Sn_5}\right)^2}{\overline{V}_{Cu}^{Cu_6Sn_5}} \Delta x^{Cu_6Sn_5}(3.43 \times 10^{-7})$$

$$\widetilde{D}_{int}^{Cu_6Sn_5} = \frac{V_m^{Cu_6Sn_5}}{\overline{V}_{Cu}^{Cu_6Sn_5}} 10.59 \times 10^{-6} \times 13 \times 10^{-6} \times (3.43 \times 10^{-7})$$

$$\widetilde{D}_{int}^{Cu_6Sn_5} = \frac{V_m^{Cu_6Sn_5}}{\overline{V}_{Cu}^{Cu_6Sn_5}} (4.72 \times 10^{-17}) \; m^2/s$$

Since the partial molar volumes are not known, if we consider that partial molar volumes are equal to the actual molar volume, we have

$$\widetilde{D}_{int}^{Cu_6Sn_5} = 4.72 \times 10^{-17} \; m^2/s$$

Following Equation S6b, we can write the interdiffusion flux with respect to component Cu as

$$\widetilde{J}_{Cu}^{Cu_6Sn_5} = \left(\frac{C_{Cu}^- - C_{Cu}^+}{2t}\right)\left[Y_{C_{Cu}}^{Cu_6Sn_5}\left(1 - Y_{C_{Cu}}^{Cu_6Sn_5}\right)\Delta x^{Cu_6Sn_5} + 0 + Y_{C_{Cu}}^{Cu_6Sn_5}\left(1 - Y_{C_{Cu}}^{Cu_3Sn}\right)\Delta x^{Cu_3Sn}\right]$$

$$\widetilde{J}_{Cu}^{Cu_6Sn_5} = \left(\frac{\frac{1}{7.12 \times 10^{-6}}}{2 \times 81 \times 3600}\right)\left[\left(\frac{6 \times 7.12}{11 \times 10.59}\right)\left(1 - \frac{6 \times 7.12}{11 \times 10.59}\right) 13 \times 10^{-6} + \left(\frac{6 \times 7.12}{11 \times 10.59}\right)\left(1 - \frac{3 \times 7.12}{4 \times 8.59}\right) 3.5 \times 10^{-6}\right]$$

$$\widetilde{J}_{Cu}^{Cu_6Sn_5} = 8.44 \times 10^{-7} \; mol/m^2.s$$

Following Equation S3,

$$\widetilde{D}_{int}^{Cu_6Sn_5} = \frac{\left(V_m^{Cu_6Sn_5}\right)^2}{\overline{V}_{Sn}^{Cu_6Sn_5}} \Delta x^{Cu_6Sn_5}\left(\widetilde{J}_{Cu}^{Cu_6Sn_5}\right)$$

$$\widetilde{D}_{int}^{Cu_6Sn_5} = \frac{\left(V_m^{Cu_6Sn_5}\right)^2}{\overline{V}_{Sn}^{Cu_6Sn_5}} \Delta x^{Cu_6Sn_5}(8.44 \times 10^{-7})$$

$$\widetilde{D}_{int}^{Cu_6Sn_5} = \frac{V_m^{Cu_6Sn_5}}{\overline{V}_{Sn}^{Cu_6Sn_5}} 10.59 \times 10^{-6} \times 13 \times 10^{-6} \times (8.44 \times 10^{-7})$$

$$\widetilde{D}_{int}^{Cu_6Sn_5} = \frac{V_m^{Cu_6Sn_5}}{\overline{V}_{Sn}^{Cu_6Sn_5}} (1.16 \times 10^{-16}) \; m^2/s$$

Again, if we consider the partial molar volumes as equal to the molar volume, we have

$$\widetilde{D}_{int}^{Cu_6Sn_5} = 1.16 \times 10^{-16} \; m^2/s.$$

This situation arises, since the relation expressed in Equation 6, *i.e.*, $\overline{V}_{Cu}^{Cu_3Sn}\widetilde{J}_{Cu}^{Cu_3Sn} + \overline{V}_{Sn}^{Cu_3Sn}\widetilde{J}_{Sn}^{Cu_3Sn} = 0$ does not fulfill with this assumption.

Therefore, we cannot consider the actual molar volume variation for the estimation of the integrated diffusion coefficient following the relations derived with respect to concentration normalized variable. Now let us examine the situations considering the ideal molar volume variations.

**S3.2 Estimation of the data considering the ideal molar volume variation**

Molar volumes of end–members: $V_m^{Cu} = 7.12 \times 10^{-6}$ and $V_m^{Sn} = 16.24 \times 10^{-6}$ m³/mol

Ideal molar volume of $\beta$ phase following the Vegard's law is $V_m^\beta = N_{Cu}^\beta V_m^{Cu} + N_{Sn}^\beta V_m^{Sn}$

Therefore, $V_m^{Cu_3Sn}(ideal) = 9.4 \times 10^{-6}$ m³/mol and $V_m^{Cu_6Sn_5}(ideal) = 11.3 \times 10^{-6}$ m³/mol

Considering the ideal variation of molar volume, we can write

$$Y_{C_{Sn}}^{Cu_3Sn} = \frac{C_{Sn}^{Cu_3Sn} - C_{Sn}^-}{C_{Sn}^+ - C_{Sn}^-} = \frac{\left(\frac{1}{4}\right)/(9.4 \times 10^{-6}) - 0}{1/(16.24 \times 10^{-6}) - 0} = \frac{1 \times 16.24}{4 \times 9.4}$$

$$Y_{C_{Sn}}^{Cu_6Sn_5} = \frac{C_{Sn}^{Cu_6Sn_5} - C_{Sn}^-}{C_{Sn}^+ - C_{Sn}^-} = \frac{\left(\frac{5}{11}\right)/(11.3 \times 10^{-6}) - 0}{1/(16.24 \times 10^{-6}) - 0} = \frac{5 \times 16.24}{11 \times 11.3}$$

$$Y_{C_{Cu}}^{Cu_3Sn} = \frac{C_{Cu}^{Cu_3Sn} - C_{Cu}^+}{C_{Cu}^- - C_{Cu}^+} = \frac{\left(\frac{3}{4}\right)/(9.4 \times 10^{-6}) - 0}{1/(7.12 \times 10^{-6}) - 0} = \frac{3 \times 7.12}{4 \times 9.4}$$

$$Y_{C_{Cu}}^{Cu_6Sn_5} = \frac{C_{Cu}^{Cu_6Sn_5} - C_{Cu}^+}{C_{Cu}^- - C_{Cu}^+} = \frac{\left(\frac{6}{11}\right)/(11.3 \times 10^{-6}) - 0}{1/(7.12 \times 10^{-6}) - 0} = \frac{6 \times 7.12}{11 \times 11.3}$$

**S3.2.1 Estimation in the Cu₃Sn phase**

We first estimate $\tilde{J}_i^\beta$ (flux of component inside $\beta$ phase) and then we see how it affects $\widetilde{D}_{int}^\beta$.

Following Equation S6a, we can write the interdiffusion flux with respect to component Sn as

$$\tilde{J}_{Sn}^{Cu_3Sn} = -\left(\frac{C_{Sn}^+ - C_{Sn}^-}{2t}\right)\left[Y_{C_{Sn}}^{Cu_3Sn}\left(1 - Y_{C_{Sn}}^{Cu_3Sn}\right)\Delta x^{Cu_3Sn} + 0 + Y_{C_{Sn}}^{Cu_3Sn}\left(1 - Y_{C_{Sn}}^{Cu_6Sn_5}\right)\Delta x^{Cu_6Sn_5}\right]$$

$$\tilde{J}_{Sn}^{Cu_3Sn} = -\left(\frac{\frac{1}{16.24 \times 10^{-6}}}{2 \times 81 \times 3600}\right)\left[\frac{16.24}{4 \times 9.4}\left(1 - \frac{16.24}{4 \times 9.4}\right)3.5 \times 10^{-6} + \frac{16.24}{4 \times 9.4}\left(1 - \frac{5 \times 16.24}{11 \times 11.3}\right)13 \times 10^{-6}\right]$$

$$\tilde{J}_{Sn}^{Cu_3Sn} = -2.96 \times 10^{-7} \text{ mol/m}^2.\text{s}$$

Following Equation S3,

$$\widetilde{D}_{int}^{Cu_3Sn} == \frac{\left(V_m^{Cu_3Sn}\right)^2}{\bar{V}_{Cu}^{Cu_3Sn}} \Delta x^{Cu_3Sn}\left(-\tilde{J}_{Sn}^{Cu_3Sn}\right)$$

$$\widetilde{D}_{int}^{Cu_3Sn} = \frac{\left(V_m^{Cu_3Sn}\right)^2}{\bar{V}_{Cu}^{Cu_3Sn}} \Delta x^{Cu_3Sn}(2.96 \times 10^{-7})$$

$$\widetilde{D}_{int}^{Cu_3Sn} = \frac{(9.4 \times 10^{-6})^2}{7.12 \times 10^{-6}} \times 3.5 \times 10^{-6} \times (2.96 \times 10^{-7})$$

$$\widetilde{D}_{int}^{Cu_3Sn} \approx 1.28 \times 10^{-17} \ m^2/s.$$

Following Equation S6b, we can write the interdiffusion flux with respect to component Cu as

$$\tilde{J}_{Cu}^{Cu_3Sn} = \left(\frac{C_{Cu}^- - C_{Cu}^+}{2t}\right) \left[Y_{C_{Cu}}^{Cu_3Sn}(1 - Y_{C_{Cu}}^{Cu_3Sn})\Delta x^{Cu_3Sn} + (1 - Y_{C_{Cu}}^{Cu_3Sn})Y_{C_{Cu}}^{Cu_6Sn_5}\Delta x^{Cu_6Sn_5} + 0\right]$$

$$\tilde{J}_{Cu}^{Cu_3Sn} = \left(\frac{\frac{1}{7.12 \times 10^{-6}}}{2 \times 81 \times 3600}\right) \left[\left(\frac{3 \times 7.12}{4 \times 9.4}\right)\left(1 - \frac{3 \times 7.12}{4 \times 9.4}\right) 3.5 \times 10^{-6} + \left(1 - \frac{3 \times 7.12}{4 \times 9.4}\right)\left(\frac{6 \times 7.12}{11 \times 11.3}\right) 13 \times 10^{-6}\right]$$

$$\tilde{J}_{Cu}^{Cu_3Sn} = 6.71 \times 10^{-7} \ mol/m^2.s$$

Following Equation S3,

$$\widetilde{D}_{int}^{Cu_3Sn} = \frac{\left(V_m^{Cu_3Sn}\right)^2}{\overline{V}_{Sn}^{Cu_3Sn}} \Delta x^{Cu_3Sn} \left(\tilde{J}_{Cu}^{Cu_3Sn}\right)$$

$$\widetilde{D}_{int}^{Cu_3Sn} = \frac{\left(V_m^{Cu_3Sn}\right)^2}{\overline{V}_{Sn}^{Cu_3Sn}} \Delta x^{Cu_3Sn} (6.71 \times 10^{-7})$$

$$\widetilde{D}_{int}^{Cu_3Sn} = \frac{(9.4 \times 10^{-6})^2}{16.24 \times 10^{-6}} \times 3.5 \times 10^{-6} \times (6.71 \times 10^{-7})$$

$$\widetilde{D}_{int}^{Cu_3Sn} \approx 1.28 \times 10^{-17} \ m^2/s.$$

Therefore, we get the same value when estimated with respect to the component Sn and Cu, since the relation expressed in Equation 6 is also fulfilled

$$\overline{V}_{Cu}^{Cu_3Sn}\tilde{J}_{Cu}^{Cu_3Sn} + \overline{V}_{Sn}^{Cu_3Sn}\tilde{J}_{Sn}^{Cu_3Sn} = 7.12 \times 10^{-6} \times 6.71 \times 10^{-7} - 16.24 \times 10^{-6} \times 2.96 \times 10^{-7} \approx 0$$

**S3.2.2 Estimation in the $Cu_6Sn_5$ phase**

Following Equation S6a, we can write the interdiffusion flux with respect to component Sn as

$$\tilde{J}_{Sn}^{Cu_6Sn_5} = -\left(\frac{C_{Sn}^+ - C_{Sn}^-}{2t}\right) \left[Y_{C_{Sn}}^{Cu_6Sn_5}(1 - Y_{C_{Sn}}^{Cu_6Sn_5})\Delta x^{Cu_6Sn_5} + (1 - Y_{C_{Sn}}^{Cu_6Sn_5})Y_{C_{Sn}}^{Cu_3Sn}\Delta x^{Cu_3Sn} + 0\right]$$

$$\tilde{J}_{Sn}^{Cu_6Sn_5} = -\left(\frac{\frac{1}{16.24 \times 10^{-6}}}{2 \times 81 \times 3600}\right) \left[\frac{5 \times 16.24}{11 \times 11.3}\left(1 - \frac{5 \times 16.24}{11 \times 11.3}\right) 13 \times 10^{-6} + \left(1 - \frac{5 \times 16.24}{11 \times 11.3}\right)\frac{16.24}{4 \times 9.4} 3.5 \times 10^{-6}\right]$$

$$\tilde{J}_{Sn}^{Cu_6Sn_5} = -3.66 \times 10^{-7} \ mol/m^2.s$$

Following Equation S3,

$$\widetilde{D}_{int}^{Cu_6Sn_5} = \frac{\left(V_m^{Cu_6Sn_5}\right)^2}{\overline{V}_{Cu}^{Cu_6Sn_5}} \Delta x^{Cu_6Sn_5} \left(-\tilde{J}_{Sn}^{Cu_6Sn_5}\right)$$

$$\widetilde{D}_{int}^{Cu_6Sn_5} = \frac{\left(V_m^{Cu_6Sn_5}\right)^2}{\overline{V}_{Cu}^{Cu_6Sn_5}} \Delta x^{Cu_6Sn_5}(3.66 \times 10^{-7})$$

$$\widetilde{D}_{int}^{Cu_6Sn_5} = \frac{(11.3 \times 10^{-6})^2}{7.12 \times 10^{-6}} \times 13 \times 10^{-6} \times (3.66 \times 10^{-7})$$

$$\widetilde{D}_{int}^{Cu_6Sn_5} \approx 8.53 \times 10^{-17} \ m^2/s$$

Following Equation S6b, we can write the interdiffusion flux with respect to component Cu as

$$\tilde{J}_{Cu}^{Cu_6Sn_5} = \left(\frac{C_{Cu}^- - C_{Cu}^+}{2t}\right)\left[Y_{C_{Cu}}^{Cu_6Sn_5}\left(1 - Y_{C_{Cu}}^{Cu_6Sn_5}\right)\Delta x^{Cu_6Sn_5} + 0 + Y_{C_{Cu}}^{Cu_6Sn_5}\left(1 - Y_{C_{Cu}}^{Cu_3Sn}\right)\Delta x^{Cu_3Sn}\right]$$

$$\tilde{J}_{Cu}^{Cu_6Sn_5} = \left(\frac{\frac{1}{7.12\times10^{-6}}}{2\times81\times3600}\right)\left[\left(\frac{6\times7.12}{11\times11.3}\right)\left(1 - \frac{6\times7.12}{11\times11.3}\right)13\times10^{-6} + \left(\frac{6\times7.12}{11\times11.3}\right)\left(1 - \frac{3\times7.12}{4\times9.4}\right)3.5\times10^{-6}\right]$$

$$\tilde{J}_{Cu}^{Cu_6Sn_5} = 8.31 \times 10^{-7} \ mol/m^2.s$$

Following Equation S3,

$$\widetilde{D}_{int}^{Cu_6Sn_5} = \frac{\left(V_m^{Cu_6Sn_5}\right)^2}{\overline{V}_{Sn}^{Cu_6Sn_5}} \Delta x^{Cu_6Sn_5}\left(\tilde{J}_{Cu}^{Cu_6Sn_5}\right)$$

$$\widetilde{D}_{int}^{Cu_6Sn_5} = \frac{\left(V_m^{Cu_6Sn_5}\right)^2}{\overline{V}_{Sn}^{Cu_6Sn_5}} \Delta x^{Cu_6Sn_5}(8.31 \times 10^{-7})$$

$$\widetilde{D}_{int}^{Cu_6Sn_5} = \frac{(11.3 \times 10^{-6})^2}{16.24 \times 10^{-6}} \times 13 \times 10^{-6} \times (8.31 \times 10^{-7})$$

$$\widetilde{D}_{int}^{Cu_6Sn_5} \approx 8.53 \times 10^{-17} \ m^2/s$$

Therefore, we have the same value when estimated with respect to the component Sn and Cu, since Equation 6 fulfills following

$$\overline{V}_{Cu}^{Cu_6Sn_5}\tilde{J}_{Cu}^{Cu_6Sn_5} + \overline{V}_{Sn}^{Cu_6Sn_5}\tilde{J}_{Sn}^{Cu_6Sn_5} = 7.12\times10^{-6} \times 8.31\times10^{-7} - 16.24\times10^{-6} \times 3.66\times10^{-7} \approx 0$$

Therefore, we can conclude that if the relation with respect to the concentration normalized variable is used for the estimation of the interdiffusion coefficient, we need to consider the ideal molar volume. We cannot consider the actual molar volumes of the phases.

*One important fact should be noted here that*

*For pure end–members, $N_B^- = 0$, $N_A^+ = 0$, $N_B^+ = 1$ and $N_A^- = 1$.*

*Therefore, we have $C_B^- = 0$, $C_A^+ = 0$, $C_B^+ = \frac{1}{V_m^+}$ and $C_A^- = \frac{1}{V_m^-}$ such that*

$$Y_{C_B} = \frac{C_B - C_B^-}{C_B^+ - C_B^-} = \frac{C_B}{C_B^+} = \frac{N_B}{V_m} \times V_m^+ = \frac{Y_{N_B}}{V_m} \times V_m^+$$

# Supplementary File

$$Y_{C_A} = \frac{C_A - C_A^+}{C_A^- - C_A^+} = \frac{C_A}{C_A^-} = \frac{N_A}{V_m} \times V_m^- = \frac{Y_{N_A}}{V_m} \times V_m^-$$

*While using the relations with respect to the composition normalized variables, only the molar volumes of phases is considered irrespective of the choice of diffusion profile of component A or B; however, the same is not true while using the relations with respect to the concentration normalized variables, since depending on the choice of diffusion profile, the molar volume ($V_m^+$ or $V_m^-$) of one of the end–members along with that of the phases is always being considered for the estimation of a particular diffusion parameter, leading to very minor difference in the estimated values.*